\documentclass[12pt]{article}
\usepackage{color}
\usepackage{epsfig}
\usepackage{amssymb,amsmath}
\newcommand{\lsim}{\,{\buildrel < \over {_\sim}}\,}
\newcommand{\gsim}{\,{\buildrel > \over {_\sim}}\,}

\newcommand{\comment}[1]{}

\begin{document}

\begin{titlepage}
\begin{flushright}
HIP-2008-03/TH\\
1 February, 2008\\
revised 1 July, 2008
\end{flushright}
\vfill
\begin{centering}

{\bf AN IMPROVED GLOBAL ANALYSIS OF NUCLEAR PARTON DISTRIBUTION FUNCTIONS
INCLUDING RHIC DATA}

\vspace{0.5cm}
Kari J. Eskola$^{\rm a,b,}$\footnote{kari.eskola@phys.jyu.fi},
Hannu Paukkunen$^{\rm a,b,}$\footnote{hannu.paukkunen@phys.jyu.fi} and 
Carlos A. Salgado$^{\rm c,}$\footnote{carlos.salgado@cern.ch}

\vspace{1cm}
{\em $^{\rm a}$Department of Physics,
P.O. Box 35, FI-40014 University of Jyv\"askyl\"a, Finland}
\vspace{0.3cm}

{\em $^{\rm b}$Helsinki Institute of Physics,
P.O. Box 64, FI-00014 University of Helsinki, Finland}
\vspace{0.3cm}

{\em $^{\rm c}$Departamento de F\'\i sica de Part\'\i culas and IGFAE, Universidade de Santiago de 
Compostela, Spain}

\vspace{1cm} 
{\bf Abstract} \\ 
\end{centering}

\noindent

We present an improved leading-order global DGLAP analysis of nuclear
parton distribution functions (nPDFs), supplementing the traditionally used
data from deep inelastic lepton-nucleus scattering and Drell-Yan dilepton
production in proton-nucleus collisions, with inclusive high-$p_T$ hadron
production data measured at RHIC in d+Au collisions.
With the help of an extended definition of the $\chi^2$ function, we now
can more efficiently exploit the constraints the different data sets offer,
for gluon shadowing in particular, and account for the overall data
normalization uncertainties during the automated $\chi^2$ minimization.
The very good simultaneous fit to the nuclear hard process data used
demonstrates the feasibility of a universal set of nPDFs, but also 
limitations become visible.
The high-$p_T$ forward-rapidity hadron data of BRAHMS add a new crucial
constraint into the analysis by offering a direct probe for the nuclear
gluon distributions -- a sector in the nPDFs which has traditionally been
very badly constrained. We obtain a strikingly stronger gluon shadowing
than what has been estimated in previous global analyses. The obtained nPDFs
are released as a parametrization called EPS08.

\vfill
\end{titlepage}

\setcounter{footnote}{0}

\section{Introduction}

With collider energies presently reached at BNL-RHIC, hard processes have become more and more 
important as diagnostic tools in the phenomenology of heavy ion collisions, QCD matter and QCD 
dynamics. The soon starting LHC heavy-ion program will emphasize the role of hard processes even 
further, by extending the kinematical range probed in the longitudinal momentum fraction $x$ and in 
the process virtuality scale $Q^2$ by several orders of magnitude with respect to the presently 
accessible ones. The existence of well-constrained up-to-date nuclear parton distribution functions 
(nPDFs) will thus be essential for the correct interpretation of the data.

Sets of collinearly factorized universal nPDFs, which are obtained in global perturbative QCD 
analyses paralleling those for the free proton, are available 
\cite{Eskola:1998iy,Eskola:1998df,Eskola:2007my,Hirai:2001np,Hirai:2004wq,deFlorian:2003qf,Hirai:2007
sx}, see also Ref. \cite{Schienbein:2007fs}. These nPDFs are typically obtained by first 
parametrizing the nuclear corrections for each parton flavour relative to a known set of the free 
proton PDFs, and imposing constraints from sum rules at a chosen initial scale $Q_0^2$. A best fit to 
nuclear hard-process data from deep inelastic lepton-nucleus scattering (DIS) and the Drell-Yan (DY) 
process in proton-nucleus collisions is obtained by an iterative procedure which involves the DGLAP 
evolution \cite{DGLAP} of the (absolute) nPDFs. The best global fit then fixes the initial nuclear 
corrections. 

Since the first of the nPDF sets, EKS98 \cite{Eskola:1998iy,Eskola:1998df}, the procedure has been 
improved by performing the analysis at next-to-leading order (NLO) 
\cite{deFlorian:2003qf,Hirai:2007sx} and by making uncertainty estimates 
\cite{Eskola:2007my,Hirai:2004wq,Hirai:2007sx} in analogy with the free proton case. In spite of such 
important progress, however, new data sets of relevance, which would more directly constrain the 
nuclear gluons in particular, have not been included in these ten years. The purpose of this paper is 
to make progress precisely in this respect, by including the data from inclusive high-$p_T$ hadron 
production in d+Au collisions at RHIC in the global analysis of the nPDFs for the first time.

The most serious difficulty in the global DGLAP analyses of nPDFs has traditionally been the lack 
of experimental data which would impose stringent enough constraints for the nuclear gluon 
distributions. The extraction of the nPDFs would be cleanest in DIS but basically no high-precision 
data are at hand in the perturbative region $Q^2\gsim 1$~GeV$^2$ at $x\lesssim 0.01$ there. 
This deficiency translates into a bad determination of the nPDFs, gluons in particular, in a region, 
where the nuclear effects are sizeable and which will be frequently accessed at the LHC. For the 
analysis of hard processes taking place at midrapidities at RHIC the situation has been better, as 
the available DIS and DY data constrain the nPDFs in most of the kinematic range probed. In the 
forward-rapidity domain, however, the hard processes are sensitive to nPDFs at smaller values of $x$ 
than what is presently constrained by DIS and DY: These RHIC data offer a possibility for independent 
further constraints of the nuclear gluon sector in the global analysis.

As discussed in previous works, \cite{Eskola:2007my, Vogt:2004hf}, the suppression in high-$p_T$ 
hadron production at forward rapidities in d+Au collisions relative to p+p collisions, measured by 
BRAHMS \cite{Arsene:2004ux}, would seem to suggest stronger gluon shadowing than, \emph{e.g.}, in the 
EKS98 set. This suppression has been also proposed as a signal of parton saturation being reached at 
RHIC \cite{satur}, so that the compatibility of the measured suppression with DIS and DY data sets 
within the DGLAP framework is a question of special relevance from the QCD parton dynamics point of 
view as well. 

In this paper, we shall demonstrate for the first time, that a global fit of a very good quality can 
indeed be obtained by simultaneously accommodating the DIS, DY and high-$p_T$ RHIC data in the 
leading-order (LO) DGLAP framework, \emph{i.e.} that a relevant new set of universal, 
process-independent, collinearly factorized nPDFs can indeed be extracted, and that the gluon 
shadowing obtained at the smallest values of $x$ is indeed stronger than in previous global fits. 
Also limitations and uncertainties remaining in the analysis 
are discussed in light of the results obtained. 
An important new feature which we introduce in the global $\chi^2$-analysis here -- borrowing it from 
the free proton analyses \cite{Stump:2001gu} -- is the treatment of data normalization errors given 
by the RHIC experiments. In particular, we demonstrate that only by accounting for these systematic 
errors, a meaningful comparison with the RHIC high-$p_T$ pion data 
\cite{Adler:2003ii,Adler:2006wg,Adams:2006nd} can be done.

Parametrizing the obtained nuclear effects for each parton flavour in $x$, $Q^2$ and $A$
and making a simple fast computer code available for public use has proven to be a working idea in 
the past. Analogously with our previous EKS98 set \cite{Eskola:1998iy,Eskola:1998df}, we now release 
a new set of nPDFS called EPS08, which is available at \cite{nPDF_address}.
 
The rest of this paper is organized as follows: In Sec.~\ref{sec:framework} we present the framework
and analysis method, introducing the functional forms used and, in particular, the improvements in 
the $\chi^2$ fitting procedure. 
In Sec.~\ref{sec:results} we present the results from the global fit, the new set of nPDFs, and the 
comparison with experimental data. The effects of the normalization-error treatment are demonstrated. 
In Sec.~\ref{sec:discussion} we comment on the strong gluon shadowing solution found. Conclusions and 
outlook are presented in Sec.~\ref{sec:conclusions}.

\section{Framework and analysis method}
\label{sec:framework}

\subsection{Definition of nPDFs}
\label{subsec:nPDFs}
The DGLAP framework we use is essentially the same as in our previous global fits of nPDFs 
\cite{Eskola:1998iy, Eskola:1998df, Eskola:2007my}. For each parton flavour $i$, we define 
the nPDFs $f_i^A(x,Q^2)$ as the PDFs of protons bound to a nucleus of mass number $A$,
\begin{equation}
f_{i}^A(x,Q^2) \equiv R_{i}^A(x,Q^2) f_{i}^{\rm CTEQ6L1}(x,Q^2),
\label{eq:nPDFdefinition}
\end{equation}  
where $f_{i}^{\rm CTEQ6L1}(x,Q^2)$ is obtained from the latest LO CTEQ set of the free proton PDFs 
\cite{Pumplin:2002vw}, and $R_{i}^A(x,Q^2)$ is the nuclear modification factor
for this parton flavour. We also assume that PDFs of bound neutrons can be obtained on the basis of  
isospin symmetry. For instance, the total $u$ quark PDF in a nucleus $A$ with $Z$ protons then 
becomes
$$
u_A(x,Q^2) = Z f_u^A(x,Q^2) + (A-Z)f_d^A(x,Q^2).
$$
The nuclear effects of Deuterium ($A=2$) and those in the cumulative region $x>1$ are neglected. 
We do not discuss the dependence of the nPDFs
on transverse location inside the nucleus (the impact parameter dependence) 
\cite{Vogt:2004hf,Eskola:1991ec} here, either, only the average nuclear effects are considered.

We parametrize the nuclear modifications $R_{i}^A$ at an initial scale $Q_0^2=1.69 \, {\rm GeV}^2$, 
which conveniently matches the lowest scale and the charm-quark mass threshold in the CTEQ6L1 set. 
The valence quark modifications are constrained by baryon number conservation and 
the gluon modifications by momentum conservation. At higher scales $Q^2 > Q^2_0$ the nPDFs 
are then obtained by solving the conventional DGLAP equations \cite{DGLAP} at LO numerically, 
applying the fast solution method introduced in \cite{Santorelli:1998yt}.

In order to reduce the amount of fitting parameters, and obtain a converging 
well-constrained fit, 
we have to assume initially flavour-independent nuclear effects for valence quarks, sea quarks and 
gluons, \emph{i.e.} we parametrize only three different functions: $R_{V}^A(x,Q_0^2)$ for all valence 
quarks, $R_{S}^A(x,Q_0^2)$ for all sea quarks, and $R_{G}^A(x,Q_0^2)$ for gluons. In the DGLAP 
evolution,
each flavour is considered individually so that at $Q^2>Q^2_0$ the modifications may in principle 
depend on parton flavour. 

\subsection{Fitting functions and parameters}
\label{subsec:parameters}

Choosing a suitable functional form for the input nuclear modification factors $R_{i}^A(x,Q_0^2)$ is 
among the most troublesome and crucial issues in the global nPDF analyses. On one hand the fit 
functions must be flexible enough, \emph{i.e.} there should appear sufficiently many parameters so 
that all the relevant features suggested by the data can be caught. On the other hand, too large a 
number of parameters easily leads to badly converging fits. Thus the number of parameters always is a 
compromise between the flexibility of the fit and the feasibility of the $\chi^2$-analysis. 

Up to now, gluon shadowing in the small-$x$ region $x \lesssim 10^{-2}$ has been constrained only by 
the momentum sum rule, and it has been essentially dictated by the assumed form of the fit function. 
Also the statistical error bars computed then reflect the uncertainties within the fit function 
chosen. Thus, as discussed in \cite{Eskola:2007my}, there has been a large uncontrolled error in the 
gluon shadowing. The inclusion of the high-$p_T$ hadron data from RHIC at forward rapidities, 
however, now provides important further constraints for the gluon shadowing region, which will be 
exploited in the present work. 

In order to take into account the RHIC high-$p_T$ forward-rapidity hadron data, which suggest a 
stronger gluon shadowing than obtained in previous global analyses (see the discussion in 
\cite{Eskola:2007my}), we need to modify the parametrization of the shadowing region. In our past 
works, 
we assumed a saturation of the modifications  $R_{i}^A(x,Q_0^2)$ such that $R_G^A\approx 
R_{S}^A\rightarrow const$ at $x\rightarrow 0$. This assumption is relaxed in the present analysis, 
and we introduce a power-law behaviour of the nuclear modification factors at small $x$ as follows
\begin{equation}
R^A_i(x,Q_0^2) \stackrel{x \rightarrow 0} {\sim} x^{\alpha^A}, \qquad  \alpha^A > 0.
\label{eq:asympt}
\end{equation}
This can be motivated by the (approximate) power-law behaviour $\sim x^{-P_{\rm free}}$ of the free 
proton PDFs at small-$x$, assuming that the nPDFs share this same gross feature but with a different 
power $P_{\rm free} \ge P_{\rm bound}$. This means that $R_i^A \sim x^{\left(P_{\rm free}-P_{\rm 
bound}\right)} \rightarrow 0$ as $x\rightarrow 0$.

\begin{figure}[h]
\center
\includegraphics[scale=0.7]{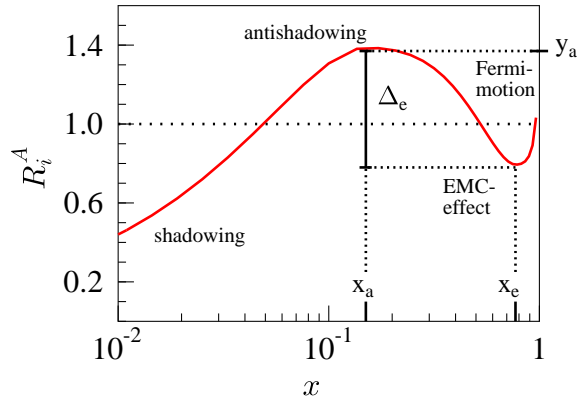}
\caption{\small An illustration of the smoothly matched fit functions $R_i^A(x)$ and the role of the 
parameters $x_a^A$, $y_a^A$, $x_e^A$ and $\Delta_e^A$. The superscripts $A$ have been 
suppressed in the figure.
}
\label{fig:RiA_illustration}
\end{figure}

With this assumption as a guide, we parametrize the initial nuclear modifications $R_V^A$, $R_S^A$ 
and $R_G^A$ in three pieces as illustrated in Fig.~\ref{fig:RiA_illustration}: $R_1^A(x)$ at small 
values of $x$, below the antishadowing maximum,  $x \le x_a^A$; $R_2^A(x)$ from the antishadowing 
maximum to the EMC minimum, $x_a^A \le x \le x_e^A$; and $R_3^A(x)$ in the  large-$x$ Fermi-motion 
region, $x\ge x_e^A$;

\begin{equation}
 \begin{array}{ll}
  R_1^A(x) = c_0^A+(c_1^A+c_2^A x^{\alpha^A})[\exp(-x/x_s^A) 
           - \exp(-x_a^A/x_s^A)],   &  x\le x_a^A \\
  R_2^A(x) = a_0^A + a_1^A x + a_2^A x^2 + a_3^A x^3, 
             & x_a^A \le x \le x_e^A \\ 
  R_3^A(x) = \frac{b_0^A-b_1^A x}{(1-x)^{\beta^A}} +  b_2^A \left(x-x_e\right)^2,      
             & x_e^A \le x \le 1.
\end{array}
\label{eq:R}
\end{equation}
Some of the parameters above are eliminated by matching the different pieces smoothly together:
we require continuity of the fit functions and zero first-derivatives at the antishadowing 
maximum $x_a^A$ and at the EMC minimum $x_e^A$. The required behavior (\ref{eq:asympt}) fixes 
$c_0^A$. It is convenient to express the fit functions in terms of the following 8 parameters, 
\\
\begin{tabular}{ll}
  $\alpha^A$             & the power according to which $R_1^A \rightarrow 0$ at $x\rightarrow 0$,\\
  $x_s^A$                & a slope factor in the exponential, \\
  $x_a^A$, $y_a^A$       & position and height of the antishadowing maximum \\
  $x_e^A$,               & position of the EMC minimum \\
  $\Delta_e^A$  		 & difference of the antishadowing maximum and the EMC minimum \\
  $\beta^A$, $b_2^A$	 & slope factors in the Fermi-motion part $R_3$ at $x>x_e$. \\
\end{tabular} 
\\

In principle, all parameters above are different from one nucleus to another, hence the superscript 
$A$. For the $A$ dependence we assume a simple power law,
\begin{equation}
  z_i^A = z_i^{A_{\rm ref}} (\frac{A}{A_{\rm ref}})^{\,p_{z_i}},
  \label{eq:Adependence}
\end{equation}
where $z_i = x_s, x_a, y_a\ldots$, and choose the reference nucleus to be Carbon, $A_{\rm ref}=12$.

With momentum and baryon number sum rules, we can fix $\alpha^A$ for gluons and valence quarks, and 
thus reduce the total number of parameters to 44. This is still far too many for a convergent 
$\chi^2$-minimization with the nuclear data constraints available. To proceed, additional assumptions 
need to be introduced -- how the unconstrained regions of the phase space are handled, and how the 
number of final fit parameters is reduced down to 15 by fixing those parameters which cannot be 
constrained, is explained in Sec.~\ref{subsec:finalparameters}.

\subsection{Data sets}
\label{subsec:Dsets}

The experimental data, providing the nonperturbative input for the nPDFs, in our present analysis 
covers three types of hard processes involving nuclei: In addition to the data from lepton-nucleus 
deep inelastic scattering and proton-nucleus Drell-Yan dilepton production, as a new ingredient we 
include the data
from inclusive high-$p_T$ hadron production in minimum-bias d+Au collisions from the BRAHMS, PHENIX 
and STAR collaborations at RHIC. In total, we have over 600 data points covering 13 nuclei from 
Helium up to Lead. Table~\ref{Table:Data} summarizes the data sets in our analysis. 

\begin{table}
\begin{center}
{\footnotesize
\begin{tabular}{lllcccc}
 Experiment & Process &  Nuclei & Data points & $\chi^2$  &Weight & Ref.\\
\hline
\hline
 SLAC E-139	        & DIS	& He(4)/D & 18 & 2.0    & 1 & \cite{Gomez:1993ri}   \\
 NMC 95, reanalysis & DIS	& He/D  	& 16 & 12.1  & 1 & \cite{Amaudruz:1995tq} \\
 \\
 NMC 95 		        & DIS	& Li(6)/D	& 15 & 30.7  & 1 & \cite{Arneodo:1995cs} \\
 \\
 SLAC E-139 	    	& DIS	& Be(9)/D & 17 & 5.5  & 1 & \cite{Gomez:1993ri}   \\
 NMC 96             & DIS	& Be(9)/C & 15 & 4.2  & 1 & \cite{Arneodo:1996rv} \\
 \\
 SLAC E-139 	    	& DIS	& C(12)/D &  7 & 3.5   & 1 & \cite{Gomez:1993ri}   \\
 NMC 95    		      & DIS	& C/D     & 15 & 10.5  & 5 & \cite{Arneodo:1995cs} \\
 NMC 95, reanalysis & DIS	& C/D   	& 16 & 17.8  & 5 & \cite{Amaudruz:1995tq} \\
 NMC 95, reanalysis & DIS	& C/Li   	& 20 & 36.4  & 1 & \cite{Amaudruz:1995tq} \\
 FNAL-E772 		      & DY	& C/D     &  9 & 8.9   & 10 & \cite{Alde:1990im}    \\
\\
 SLAC E-139		& DIS	& Al(27)/D      & 17 & 3.6   & 1 & \cite{Gomez:1993ri}   \\
 NMC 96    		& DIS 	& Al/C        & 15 & 6.7   & 1 & \cite{Arneodo:1996rv} \\
\\
 SLAC E-139		& DIS	& Ca(40)/D      &  7 & 1.3  & 1 & \cite{Gomez:1993ri}   \\
 FNAL-E772 		& DY	& Ca/D          &  9 & 5.0  & 10 & \cite{Alde:1990im}    \\
 NMC 95, reanalysis 	& DIS	& Ca/D 		& 15 & 27.9  & 1 & \cite{Amaudruz:1995tq} \\
 NMC 95, reanalysis 	& DIS	& Ca/Li 	& 20 & 26.1  & 1 & \cite{Amaudruz:1995tq} \\
 NMC 96    		& DIS	& Ca/C          & 15 & 6.3  & 1 & \cite{Arneodo:1996rv} \\
\\
 SLAC E-139		& DIS	& Fe(56)/D      & 23 & 16.5  & 1 & \cite{Gomez:1993ri}   \\
 FNAL-E772 		& DY	& Fe/D          &  9 & 5.0  & 10 & \cite{Alde:1990im}    \\
 NMC 96     		& DIS	& Fe/C          & 15 & 11.9  & 1 & \cite{Arneodo:1996rv} \\
 FNAL-E866		& DY	& Fe/Be         & 28 & 21.6  & 1 & \cite{Vasilev:1999fa} \\
\\ 
 CERN  EMC		& DIS	& Cu(64)/D      & 19 & 12.3  & 1 & \cite{Ashman:1992kv}   \\
\\
 SLAC E-139		& DIS	& Ag(108)/D     &  7 & 2.3  & 1 & \cite{Gomez:1993ri}   \\
\\
 NMC 96    		& DIS  	& Sn(117)/C     & 15 & 10.9  & 1 & \cite{Arneodo:1996rv} \\
 NMC 96, $Q^2$ dep. $x \le 0.025$   	& DIS	& Sn/C  & 24 & 9.4  & 10 & \cite{Arneodo:1996ru} \\
 NMC 96, $Q^2$ dep. $x > 0.025$   	& DIS	& Sn/C  & 120 & 75.2  & 1 & \cite{Arneodo:1996ru} \\
\\
 FNAL-E772 		& DY	& W(184)/D      &  9 & 10.0  & 10 & \cite{Alde:1990im}    \\
 FNAL-E866		& DY	& W/Be          & 28 & 26.5  & 1 & \cite{Vasilev:1999fa} \\
\\
 SLAC E-139		& DIS	& Au(197)/D     & 18 & 6.1  & 1 & \cite{Gomez:1993ri}   \\
 RHIC-BRAHMS    		& $h^-$ prod.	& dAu/pp        & 6 & 2.2  & 40 & \cite{Arsene:2004ux} \\
 RHIC-PHENIX    		& $\pi^0$ prod.	& dAu/pp        & 35 & 21.3  & 1 & 
\cite{Adler:2003ii,Adler:2006wg} \\
 RHIC-STAR     		& $\pi^++\pi^-$ prod.	& dAu/pp        & 10 & 3.5 & 1 & \cite{Adams:2006nd} \\

\\
 NMC 96    		& DIS	& Pb/C          & 15 & 5.1 & 1 & \cite{Arneodo:1996rv} \\
\\

 \hline		   
 total 		     &        &     &  627 & 448 & &                       \\
\end{tabular}
}
\caption[]{\small The data used in this analysis. The mass numbers are indicated in parentheses and 
the number of data points refers to those falling within our kinematical cuts,
$Q^2, M^2 \ge 1.69 \, {\rm GeV}^2$ for DIS and DY, and $p_T \ge 2 \, {\rm GeV}$ for hadron production 
at RHIC. 
The quoted $\chi^2$ values correspond to the unweighted contributions of each data set.}
\label{Table:Data}
\end{center}
\end{table}

The (minimum bias)  DIS and DY data that we utilize are available as ratios of differential cross 
sections between a nucleus $A$ and a reference nucleus. We denote the cross section ratios computed 
against deuterium as 
\begin{equation}
R_{\rm DIS}^{\rm A}(x,Q^2) \equiv \frac{\frac{1}{A}d\sigma_{\rm DIS}^{l \rm 
{A}}/dQ^2dx}{\frac{1}{2}d\sigma_{\rm DIS}^{l{\mathrm d}}/dQ^2dx} \stackrel{\rm LO}{=} R_{F_2}^{\rm 
A}(x,Q^2), \hspace{1cm}
R_{\rm DY}^{\rm A}(x,M^2) \equiv \frac{\frac{1}{A}d\sigma^{\rm pA}_{\rm 
DY}/dM^2dx}{\frac{1}{2}d\sigma^{\rm pd}_{\rm DY}/dM^2dx},
\label{RF2RDY}
\end{equation}
where $x$ refers to the momentum fraction and $Q^2$ to the photon virtuality for DIS, and 
$M^2$ to the invariant mass of the lepton pair and $x$ to either $x_1$ or $x_2$ for DY. 
The scales $Q^2$ and $M^2$ also define our factorization scales, and we consider only those data 
points which lie above our initial scale: $M^2, Q^2 > 1.69 \, {\rm GeV}^2$.
	
The inclusive hadron production data at RHIC comes as the nuclear modification factor $R_{\rm dAu}$, 
the ratio between the invariant yields in d+Au and p+p collisions,  
\begin{equation}
R_{\rm dAu} = \frac{1}{\langle N_{\rm coll}\rangle} \frac{d^2 N^{\rm dAu}/dp_T d\eta}{d^2 N^{\rm 
pp}/dp_T d\eta} \stackrel{\rm min. bias}{=} \frac{\frac{1}{2A} d^2\sigma^{\rm dAu}/dp_T 
d\eta}{d^2\sigma^{\rm pp}/dp_T d\eta},
\label{R_dAu}
\end{equation}
where $p_T$ and $\eta$ denote the transverse momentum and the pseudorapidity of the observed hadron,
and $\langle N_{\rm coll}\rangle$ is the estimated average number of inelastic binary nucleon-nucleon 
collisions in a centrality class studied. The last equality holds for the minimum bias case which we 
are interested in here.
According to the QCD factorization theorem, the inclusive hadron production cross sections can be 
computed as
\begin{equation}
d\sigma^{AB\rightarrow h+X} = \sum_{ijkl} 
f_i^A (Q^2) \otimes f_j^B(Q^2) \otimes \sigma^{ij\rightarrow kl}(Q^2)
\otimes D_{k\rightarrow h+X}(Q^2_f),
\label{Eq:pQCDcrosssection}
\end{equation} 
where $f_i^A (Q^2)$ are the input nPDFs, $\sigma^{ij\rightarrow kl}(Q^2)$ is the pQCD matrix element 
squared, and $D_{k\rightarrow h+X}(Q^2_f)$ denotes the fragmentation functions for which we use the 
KKP parametrization \cite{Kniehl:2000fe}. 
Our choice for the fragmentation scale is the hadronic transverse momentum, $Q_f^2 = p_T^2$, and for 
the factorization and renormalization scale $Q^2$ we take the corresponding partonic transverse 
momentum. Notice here that all the scale choices above are simplifications in the sense that they 
could be left as additional fit parameters, and that
 the KKP fragmentation functions do not distinguish negatively and positively charged hadrons.
Such details, however, are beyond the scope of the present analysis. A more detailed discussion of 
how the needed differential cross sections (\ref{Eq:pQCDcrosssection}) are calculated in practice in 
LO can be found in \cite{Eskola:2002kv}.

We should emphasize that the following choices are made with the inclusion of the RHIC data:
\begin{itemize}

\item Since we do not discuss the impact parameter dependent nuclear effects here, 
we consider only minimum bias data for $R_{\rm dAu}$. 

\item From PHENIX and STAR, we systematically include only the pion production data. This is because 
the 'Cronin-type' enhancement seems to be much larger for baryons than for mesons 
\cite{Adler:2003ii},
signalling of the fact that still at $\sqrt{s}_{\rm NN} = 200 \, {\rm GeV}$ there might be a 
significant component of nonperturbative baryon-number transport from the beam particles. We do, 
however, include the BRAHMS data for negatively charged hadrons with the assumption that the 
antiproton content in this data sample is negligible.

\item Perturbative QCD calculations of inclusive hadron production, which are  performed strictly in 
the framework of collinear factorization (without any intrinsic transverse momentum), show similar 
features both in LO \cite{Eskola:2002kv} and in NLO \cite{Borzumati:1995ib,Aurenche:1999nz}: Below 
$p_T\sim$ a few GeV the computed cross sections for $p+p(\overline p)$ collisions start to overshoot 
the data. Hence, since we should not push the nuclear case too far either, we choose to include only 
the region $p_T\ge 2$~GeV of the RHIC data on inclusive hadron production in this analysis. More 
discussion on this choice will follow later.

\end{itemize}

\subsection{Modified $\chi^2$}

The established way of fitting a set of parameters $\{z\}$ of the PDFs against a large number of 
experimental data, is the minimization of the global $\chi^2$ function. In its simplest form the 
global $\chi^2$, the goodness parameter of the fit obtained, is defined by
\begin{equation}
\chi^2(\{z\})   \equiv  \sum _N \chi^2_N(\{z\}),
\end{equation}
where $N$ labels the experimental data sets, and
\begin{equation}
\chi^2_N(\{z\})  \equiv  \sum_{i \in N} \left[\frac{D_i - T_i(\{z\})}{\sigma_i}\right]^2,
\end{equation}
where $D_i$, $\sigma_i$, and $T_i(\{z\})$ denote the value of a single data point, its measurement 
uncertainty, and the corresponding theoretical value which depends on the parameters $\{z\}$ of PDFs.

For the nPDF analyses involving only DIS and DY data, the simple form of $\chi^2$ above has been 
sufficient, but for our current purposes a more general definition for the $\chi^2$, introduced in 
\cite{Stump:2001gu}, is needed: 
\begin{eqnarray}
\chi^2(\{z\})   & \equiv & \sum _N w_N \, \chi^2_N(\{z\}) 
\label{eq:chi2mod_1}
\\	
\chi^2_N(\{z\}) & \equiv & \left( \frac{1-f_N}{\sigma_N^{\rm norm}} \right)^2 + \sum_{i \in N} 
\left[\frac{ f_N D_i - T_i(\{z\})}{\sigma_i}\right]^2,
\label{eq:chi2mod_2}
\end{eqnarray}
where 
$w_N$ is a weight factor chosen separately for each data set,
$\sigma_N^{\rm norm}$ is the relative uncertainty in the overall normalization reported by the 
experiment, and
$f_N$ is the optimized value of the overall normalization for the data set, corresponding to each 
parameter set $\{z\}$. The reasons for the necessity of such redefinition are the following:

\begin{enumerate}
\item
By making the weight factor $w_N$ larger than 1, we can emphasize by hand the importance of those 
data sets which contain definite physics content -- such as constraints for small-$x$ gluons -- but 
whose number of data points is small. With a default value $w_N=1$, such data sets would have a 
negligible contribution to the overall $\chi^2$ and the valuable constraints they offer 
would escape unnoticed.
 
\item
In addition to the point-to-point statistical and systematic errors,
certain data sets have a significant common normalization uncertainty $\sigma_N^{\rm norm}$ for all 
data points within the set. Even if this normalization uncertainty is large, the \emph{shape} of the 
distribution formed by the data points may be a valuable constraint for the nPDFs. 
We introduce for each data set a normalization factor $f_N 
\in [1-\sigma_N^{\rm norm},1+\sigma_N^{\rm norm}]$ 
which multiplies all the experimental values within the set $N$. In connection with $f_N$, there is 
an additional ``penalty'' factor $(\frac{1-f_N}{\sigma_N^{\rm norm}})^2$ which is the larger the more 
$f_N$ deviates from unity --- this accounts for the fact that having $f_N=1$ is anyway the 
experiment's best estimate for normalization. The actual value for $f_N$ is determined from the 
requirement that $\chi^2_N(\{z\})$ for each data set is at minimum.

\end{enumerate}

The motivation for both modifications in the $\chi^2$ definition discussed above comes mainly from 
adding the RHIC data for the nuclear modification factor $R_{\rm dAu}$ of Eq.~(\ref{R_dAu}) into the 
analysis. First, the BRAHMS data set for forward direction ($\eta \sim 2...3$, especially with our 
choice $p_T\ge 2$~GeV) has only a very few data points. These would not have much effect in the 
global $\chi^2$ without being artificially emphasized. Second, the average number of inelastic binary 
nucleon-nucleon collisions $\langle N_{\rm coll}\rangle$ in $\rm d+Au$ collision is derived from a 
simulation of the experiment with the Glauber model as an input, which gives rise to a significant 
model-dependent normalization uncertainty in $R_{\rm dAu}$. 

The amount of DIS data overwhelms that of the DY data. To improve upon this balance, we weight  the 
FNAL-E772 DY data set by $w_N=10$. This improves the determination of the relative importance between 
the valence and sea quarks at intermediate values of $x$.  For the NMC data set for $R_{F_2}^C$ we 
give a weight $w_N=5$ in order to better ensure a good fit for the Carbon nucleus, which is used as 
reference in the analysis. The NMC 96 data on the $Q^2$-dependence of $F_2^{\rm Sn}/F_2^{\rm C}$ is 
weighted by $w_N=10$ but only for the three lowest values of $x$ -- the three upper panels in Fig. 
\ref{Fig:RF2SnC} below -- to help constraining the gluon distribution at $x\lesssim 0.02$ via the 
DGLAP evolution. Finally, the few points of the BRAHMS data that we include in our analysis, are 
weighted by a large factor $w_N=40$ in order to account for the constraints this data set gives for 
the gluon distribution.  All these weights are summarized in Table \ref{Table:Data} above. 

\section{Results}
\label{sec:results}

\subsection{Final parameters}
\label{subsec:finalparameters}

In order to reach a well converging (well constrained) global fit, where none of the fit parameters
are drifting to their limits, we are forced to reduce the total number of free parameters down to the 
following 15: 

\begin{itemize}
 \item {\bf Valence quark modification} \\
The DIS data constrain the modification $R_V^A(x,Q_0^2)$ in the $x \gtrsim 0.1$ region rather well, 
and altogether $8$ parameters $x_a$, $y_a$, $p_{y_a}$, $x_e$, $\Delta_e$, $p_{\Delta_e}$, $b_2$, 
$p_{b_2}$ were left free. 

 \item {\bf Sea quark modification} \\
The DIS and DY data probe the sea quarks in the region $0.01 \lesssim x \lesssim 0.1$, and $5$ 
parameters,  $\alpha$, $p_{\alpha}$, $x_a$, $y_a$, $p_{y_a}$, controlling this region in 
$R_S^A(x,Q_0^2)$, were left free. The region $x\gsim 0.3$ is, however, not constrained by any present 
experimental data. We assume for simplicity a smooth behavior in this region, without an EMC effect.

 \item {\bf Gluon modification} \\
The gluon modification $R_G^A(x,Q_0^2)$ at small $x$ is now directly constrained by the inclusive 
hadron production data from RHIC. Indirectly the gluons are constrained by the $Q^2$-evolution 
effects in the sea quark sector, reflected by the DIS and DY data. In spite of the new constraints, 
we were still able to leave only $2$ parameters, $y_a$ and $p_{y_a}$ controlling the antishadowing 
peak height, free. 
We assume a similar EMC-effect for gluons as there is for valence quarks, guided by the shape of the 
preliminary PHENIX data for inclusive photon production in Au+Au collisions at $p_T\gsim 6$~GeV 
\cite{Tannenbaum:2007sy} and also by the PHENIX data 
for inclusive pion production at $\eta=0$ and $p_T\gsim 6$~GeV \cite{Adler:2006wg}
(see Fig.~\ref{Fig:PHENIX_STAR} ahead -- we have checked that indeed some sensitivity to gluon PDFs 
persists even at these values of $p_T$). 

\end{itemize}

\begin{figure}[h]
\center
\includegraphics[scale=0.7]{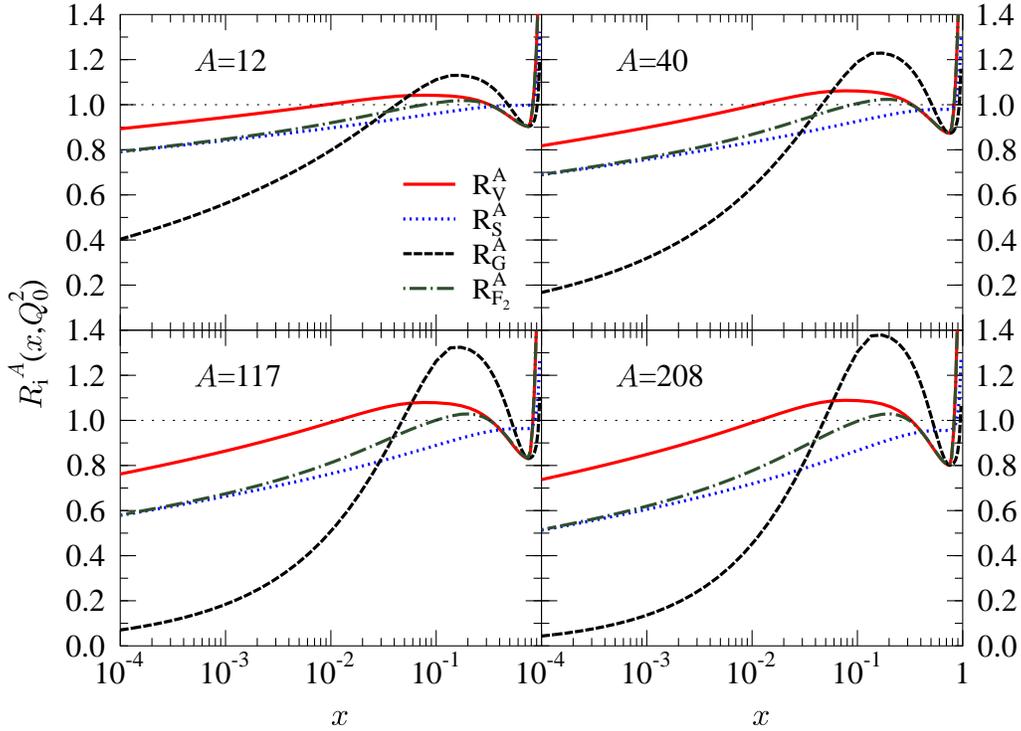}
\caption{\small The nuclear modification factors $R_V^A$, $R_S^A$ and $R_G^A$ for C, Ca, Sn, and Pb 
at $Q_0^2 = 1.69 \, {\rm GeV}^2$. The DIS ratio $R_{F_2}^A$ is shown for comparison.}
\label{fig:RAinit}
\end{figure}

The global fit with these 15 parameters was then performed by minimizing the $\chi^2$ function 
defined in Eqs.~(\ref{eq:chi2mod_1}) and (\ref{eq:chi2mod_2}) with the MINUIT \cite{James:1975dr} 
routine from the CERN Program Library. Table \ref{Table:Params} summarizes the parameter values 
obtained as well as the fixed parameters.  The goodness of the fit is characterized by $\chi^2/N = 
0.71$, where $\chi^2$ is computed with no extra weights, 
$w_N=1$, but with the optimized normalization factors $f_N$ included, and $N=627$ is the total number 
of data points. 
The contribution from each data set to this $\chi^2$ can be read off from Table~\ref{Table:Data}.
The corresponding nuclear modifications for selected nuclei at our initial scale $Q_0^2=1.69 \, {\rm 
GeV}^2$ are shown in Fig.~\ref{fig:RAinit}. 
Table~\ref{Table:Compare_to_EKPS} shows the contribution from different types of hard processes to 
the unweighted $\chi^2$ and the comparison with our previous analysis in Ref.~\cite{Eskola:2007my}. 
The $\chi^2/N$ obtained in earlier global analyses can be found in Table~3 of Ref. 
\cite{Eskola:2007my}.
\vspace{0.5cm}
 
\begin{table}[htb]
\begin{center}
\begin{tabular}{l|c|lll}
   & Param.   	&  Valence  $R_V^A$    	&  Sea $R_S^A$          	&  Gluon $R_G^A$\\
\hline
\hline
 1 &  $\alpha$ 	   & {$\alpha^A$ from baryon sum}	  &  \textbf{2.67 $\times {\bf 10^{-2}}$}       	
&  $\alpha^A$ from {momentum sum}	\\
 2 &  $p_{\alpha}$ & 	---		 	  &  \textbf{3.47 $\times {\bf 10^{-1}}$}  	&   ---	\\ 
 3 &  $x_s$    	   &  0.1, {fixed}	  &  1.0, {fixed} 	&  1.0, {fixed}    \\
 4 &  $p_{x_s}$    &  0, {fixed}   &  0, {fixed}  &  0, {fixed}   \\
 5 &  $x_a$    	   &  \textbf{7.37 $\times {\bf 10^{-2}}$} &  \textbf{0.580}        &  0.15 {fixed} 
\\
 6 &  $p_{x_a}$    &  0, {fixed}  &  0, {fixed}  &  0, {fixed}   \\
 7 &  $x_e$    	   &  \textbf{0.751}         & {as valence}  &  {as valence} \\
 8 &  $p_{x_e}$    &  0, {fixed}    &  0, {fixed}   &  0, {fixed}   \\
 9 &  $y_a$    	   &  \textbf{1.04}        	&  \textbf{0.997}       	&  \textbf{1.13}     \\
 10&  $p_{y_a}$    &  \textbf{1.55 $\times {\bf 10^{-2}}$}   	& \textbf{-1.51 $\times {\bf 
10^{-2}}$}   	&  \textbf{6.99 $\times {\bf 10^{-2}}$} \\
 11&  $\Delta_e$   &  \textbf{0.138}       	&  0, {fixed}   &  {from valence} \\
 12&  $p_{\Delta_e}$    & \textbf{0.257}    &  0, {fixed}  &  {from valence} \\
 13&  $b_2$  	     &  \textbf{13.3}          	& 0, {fixed}      &  0, {fixed} \\
 14&  $p_{b_2}$	   &  \textbf{0.278}          	& 0, {fixed}      &  0, {fixed} \\
 15&  $\beta$  	   &  0.3, {fixed} 	& 0.3, {fixed}     &  0.3, {fixed} \\
 16&  $p_{\beta}$  &  0, {fixed}   & 0, {fixed}     &  0, {fixed}	\\
\hline
\end{tabular}                                                  \\
\caption[]{\small List of all parameters defining the modifications $R_V^A$, $R_S^A$ and $R_G^A$ 
through Eq.~(\ref{eq:R}) at our initial scale $Q_0^2=1.69$~GeV$^2$. The parameters $\alpha$, $x_s$, 
$x_a$, $x_e$, $y_a$, $\Delta_e$,  $b_2$ and $\beta$ are for the reference nucleus $A=12$, and the 
powers $p_i$  define their $A$-dependence as in  Eq.~(\ref{eq:Adependence}). For valence quarks and 
gluons, the baryon number and momentum sum rules fix the parameters $\alpha^A$ for each nucleus $A$ 
separately, in which case the powers $p_{\alpha}$ are not used. The location and height of the 
EMC minimum of $R_G^A$ was fixed to that of $R_V^A$. The parameters left free for the minimization 
procedure are shown in bold face.
}
\label{Table:Params}
\end{center}
\end{table}

\begin{table}
\begin{center}
{\small
\begin{tabular}{lccc}
Data type & Data points & $\chi^2_{\rm EPS08}$ & $\chi^2_{\rm EKPS}$\\
\hline
\hline
Deep Inelastic & 484 & 344.2 & 337.4 \\
Drell-Yan & 92 & 77.1 & 84.3 \\
Hadron production & 51 & 26.9 & 28.0 \\
\hline
\hline
Total  & 627 & 448.3 & 449.6
\end{tabular}
}
\caption[]{
\small Contributions of various data types to the total unweighted 
$\chi^2$ in our previous work \cite{Eskola:2007my} (EKPS) and in this work (EPS08).}
\label{Table:Compare_to_EKPS}
\end{center}
\end{table}

\subsection{Comparison with data}
\label{subsec:compdata}

\begin{figure}[htbp]
\center
\includegraphics[scale=0.7]{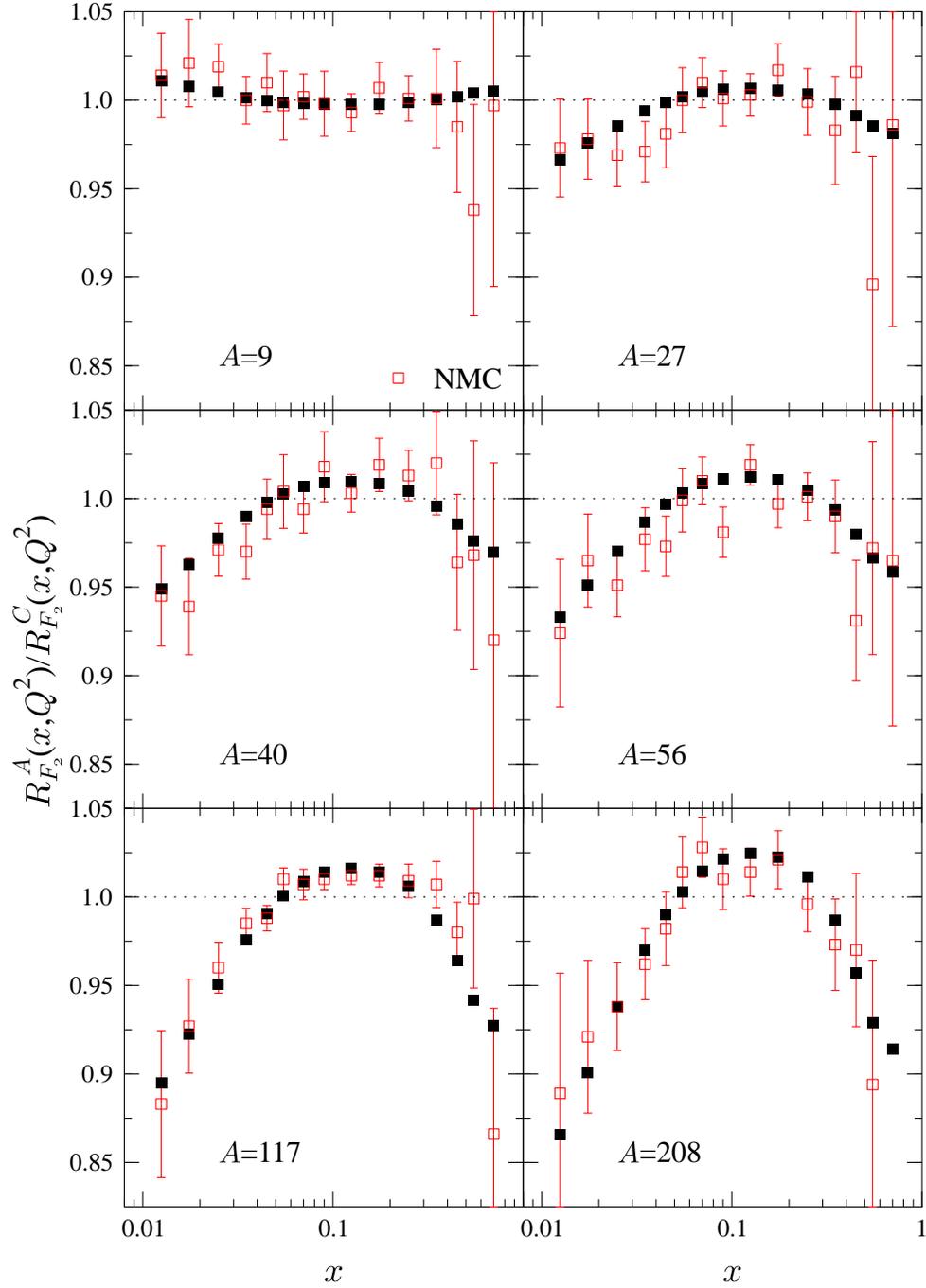}
\caption[]{\small The computed ratio $R_{F_2}^A(x,Q^2)$ vs. $R_{F_2}^{\mathrm
    C}(x,Q^2)$ compared with the NMC data \cite{Arneodo:1996rv}. The open
  symbols are the data points with statistical and systematic errors added in quadrature, the filled
  ones are the corresponding results from this analysis.}
\label{Fig:RF2AC}
\end{figure}

\begin{figure}[htbp]
\center
\includegraphics[scale=0.5]{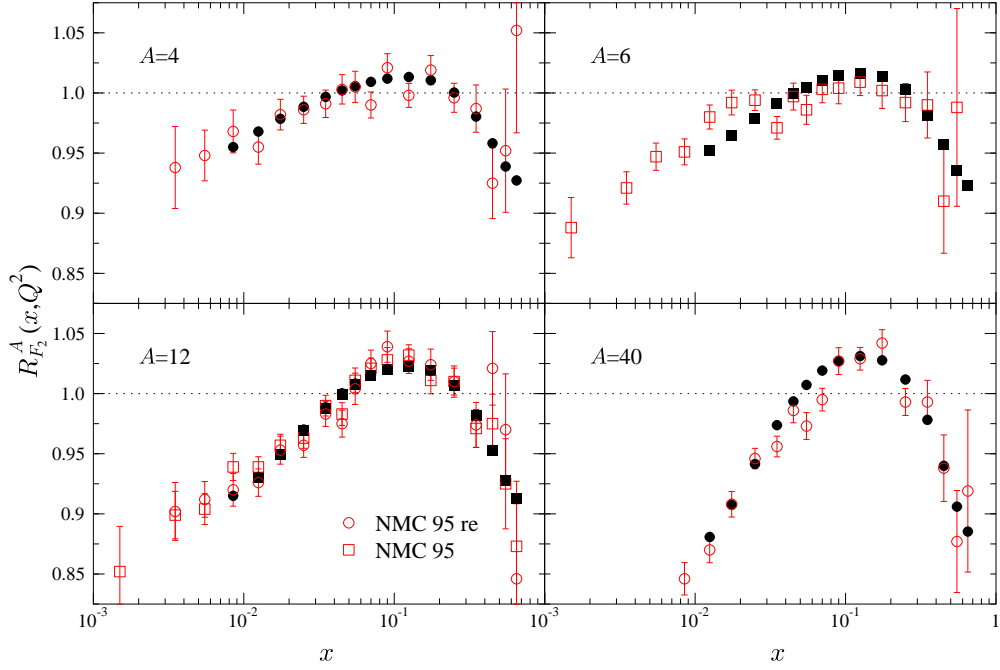}
\caption[]{\small The calculated ratio $R_{F_2}^A(x,Q^2)$  compared with the NMC 95 (squares) 
\cite{Arneodo:1995cs} and the reanalysed NMC 95 (circles) data \cite{Amaudruz:1995tq}.}
\label{Fig:RF2A1}
\end{figure}

\begin{figure}[htbp]
\center
\includegraphics[scale=0.5]{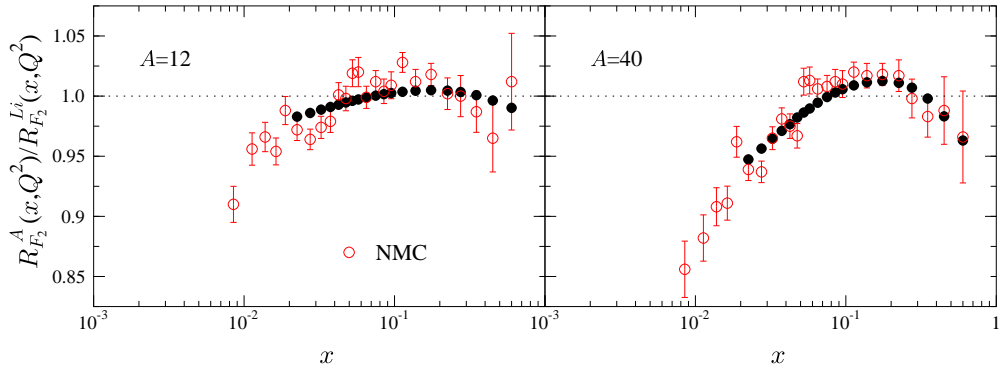}
\caption[]{\small The computed ratio $R_{F_2}^A(x,Q^2)$ vs. $R_{F_2}^{\mathrm
    Li}(x,Q^2)$ (filled circles) compared with the NMC data \cite{Amaudruz:1995tq} (open circles).}
\label{Fig:RF2ALi}
\end{figure}

\begin{figure}[htbp]
\center
\includegraphics[scale=0.45]{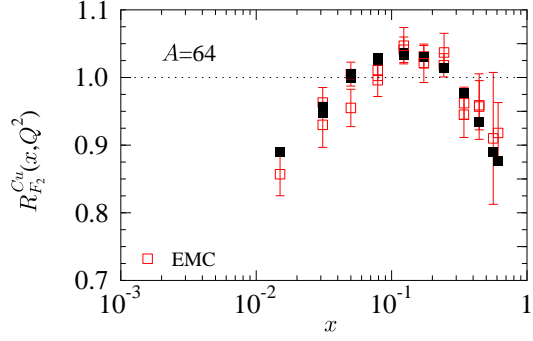} \hspace{0.5cm}
\caption[]{\small The calculated ratio $R_{F_2}^{\rm Cu}(x,Q^2)$ (filled squares) compared with the  
EMC \cite{Ashman:1992kv} data (open squares).}
\label{Fig:RF2Cu}
\end{figure}

\begin{figure}[htbp]
\center
\vspace{-0.5cm}
\includegraphics[scale=0.50]{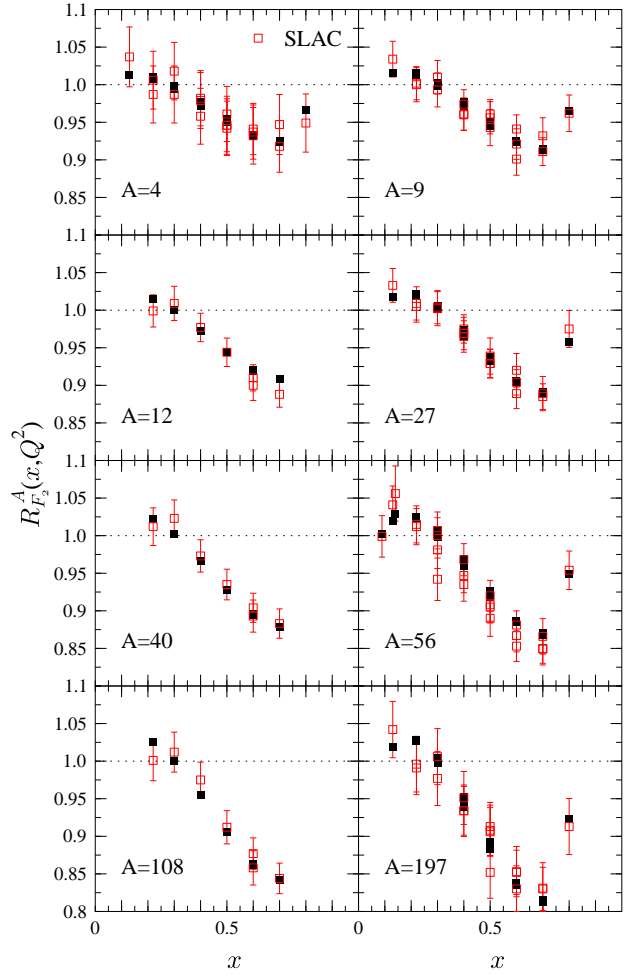}
\caption[]{\small  
The calculated ratios $R_{F_2}^A(x,Q^2)$ (filled squares) for several nuclei compared with the  
  SLAC data \cite{Gomez:1993ri} (open squares).}
\label{Fig:RF2SLAC}
\end{figure}

\begin{figure}[htbp]
\center
\includegraphics[scale=0.7]{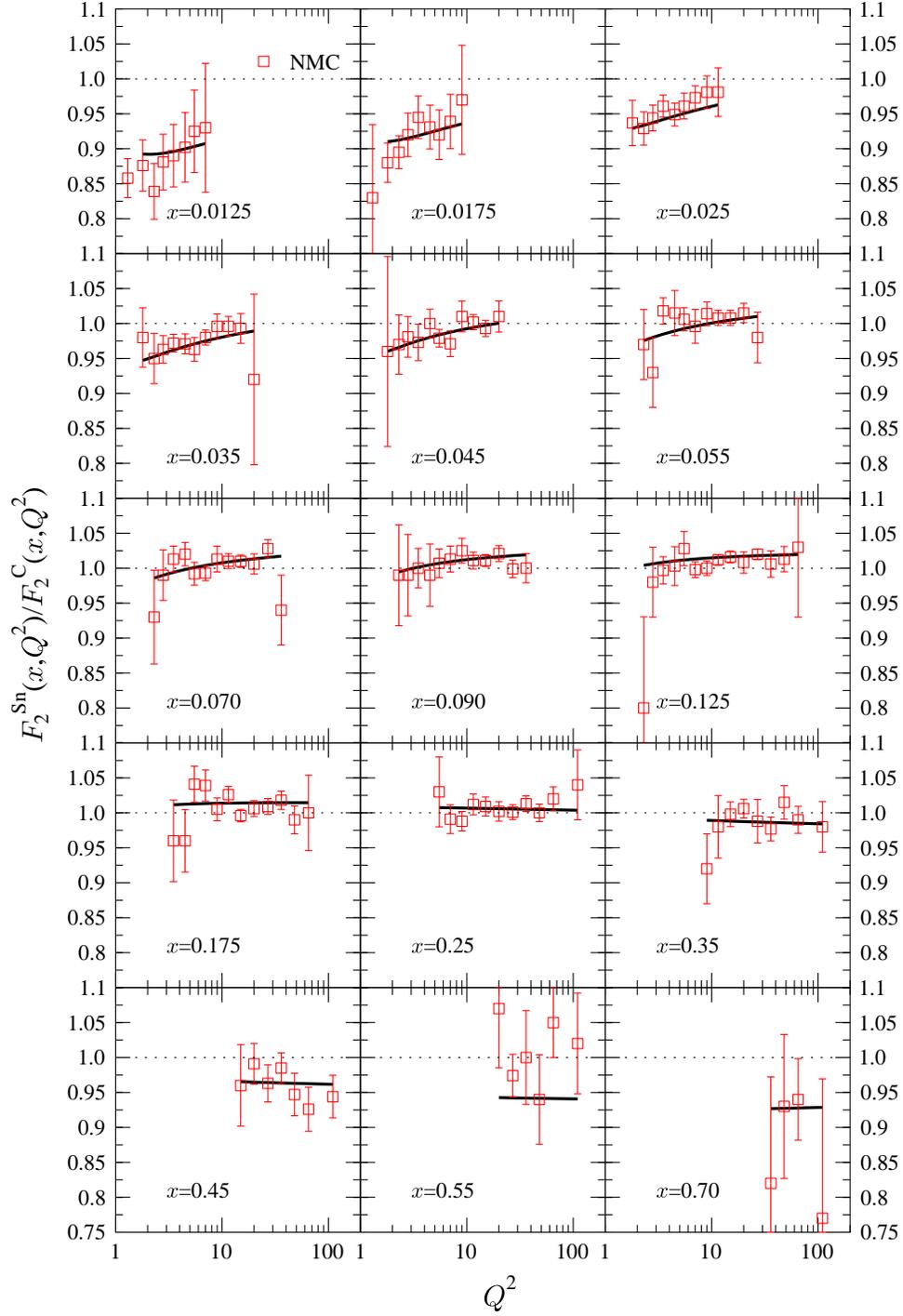}
\caption[]{\small The calculated scale evolution (solid black lines) of the ratio
  $F_2^{\mathrm{Sn}}/F_2^{\mathrm{C}}$ compared with the NMC data
  \cite{Arneodo:1996ru} for  several fixed values of $x$.}
\label{Fig:RF2SnC}
\end{figure}

\begin{figure}[htbp]
\center
\includegraphics[scale=0.45]{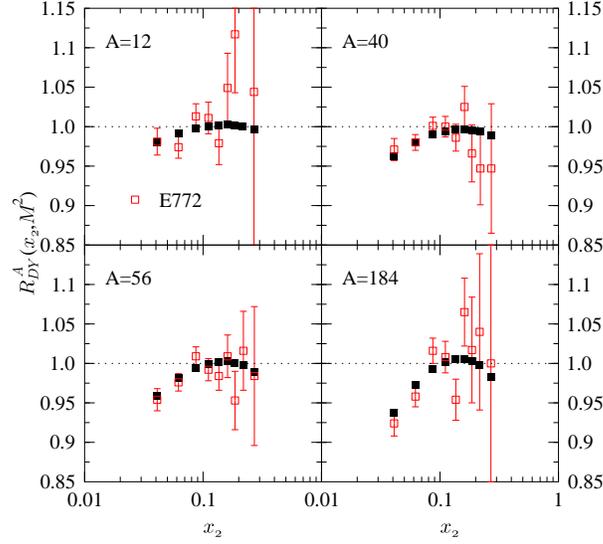}\hspace{1cm}
\vspace{-0.5cm}
\caption[]{\small The computed $R_{\rm DY}^{\rm A}(x_2,M^2)$ (filled squares) as a function of $x_2$ 
compared with the E772 data \cite{Alde:1990im} (open squares).}
\label{Fig:E772}
\end{figure}

\begin{figure}[htbp]
\vspace{-0.5cm}
\center
\includegraphics[scale=0.5]{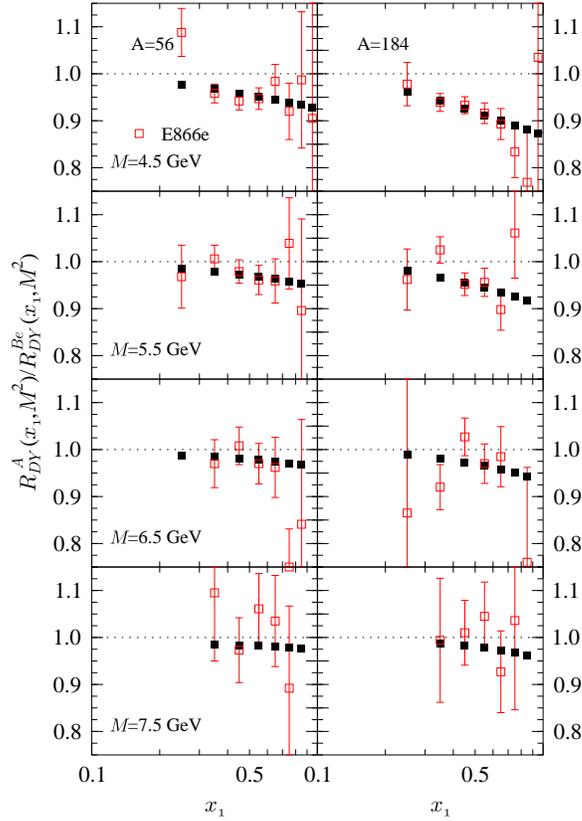}
\vspace{-0.5cm}
\caption[]{\small The computed $R_{\rm DY}^{\rm A}(x_1,M^2)$ (filled squares) as a function of $x_1$ 
compared with the E866 data \cite{Vasilev:1999fa} (open squares) at four different bins of invariant 
mass $M^2$.
}
\label{Fig:E886}
\end{figure}

\begin{figure}[htbp]
\centering
\includegraphics[width=35pc]{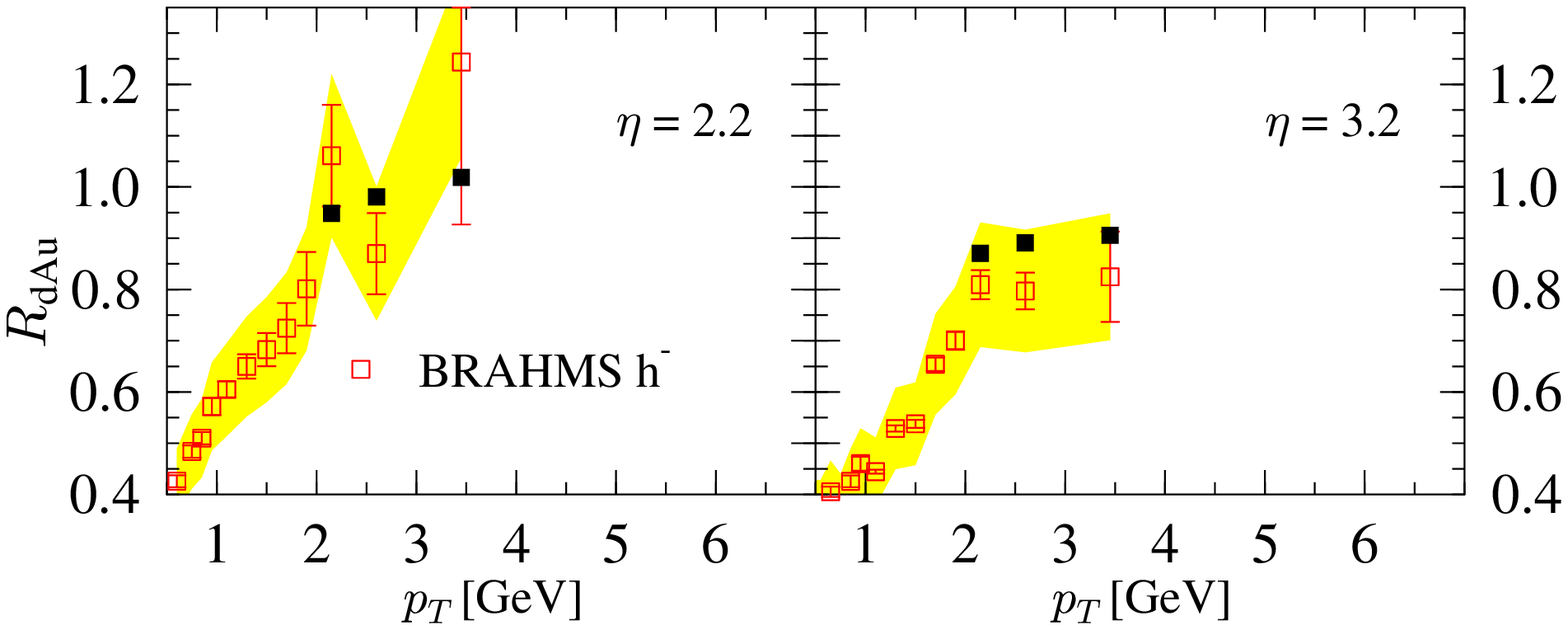}
\vspace{1cm}
\includegraphics[width=35pc]{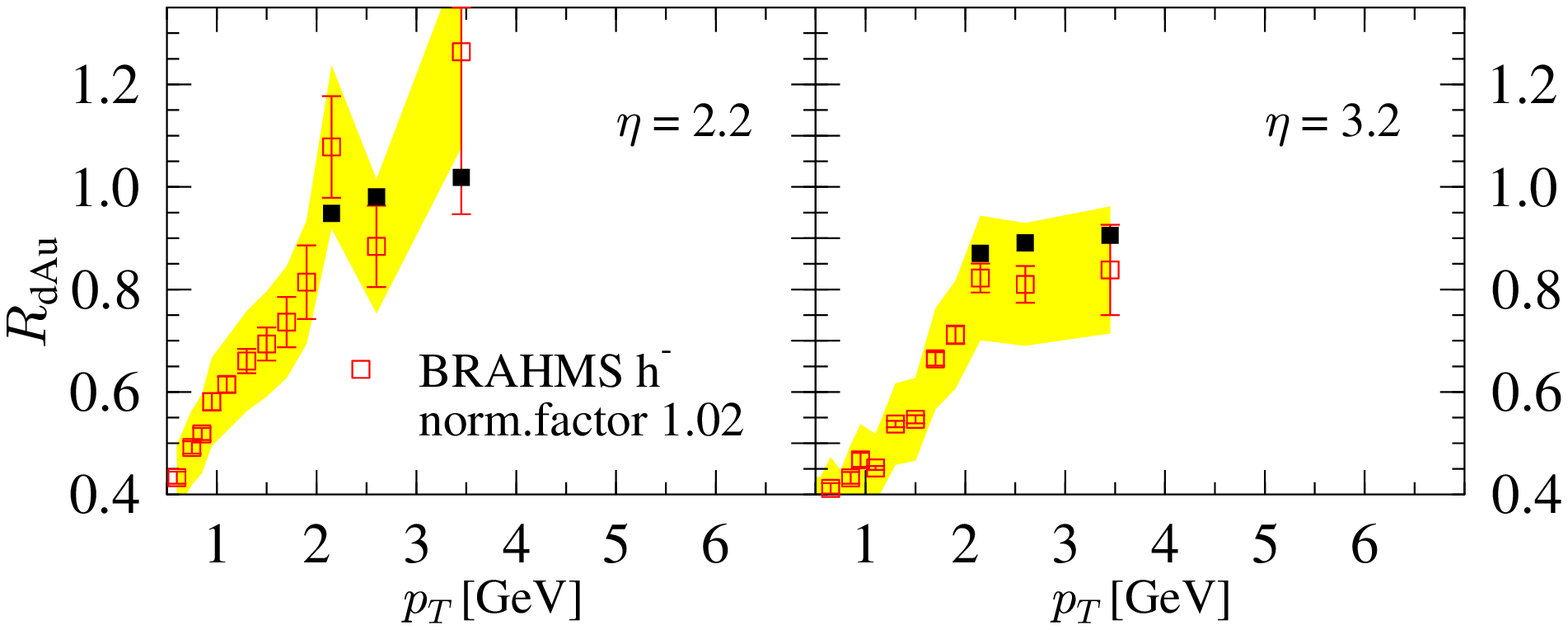}
\caption[]{\small The computed nuclear modification ratio $R_{\rm dAu}$ at forward rapidities (filled 
squares) for negatively-charged hadron production, compared with the BRAHMS  data 
\cite{Arsene:2004ux} (open squares). The error bars are the statistical uncertainties, and the shaded 
bands indicate the point-to-point systematic errors. The additional overall normalization 
uncertainty, is $5\%$, \emph{i.e.} $\sigma^{\rm norm}_N=0.05$ in Eq.~(\ref{eq:chi2mod_2}). The upper 
panels show the comparison without the normalization factor $f_N$. In the lower panels, the data have 
been multiplied by the optimized value $f_N = 1.02$.
}
\label{Fig:BRAHMS}
\end{figure}

\begin{figure}[htbp]
\centering
\includegraphics[width=18pc]{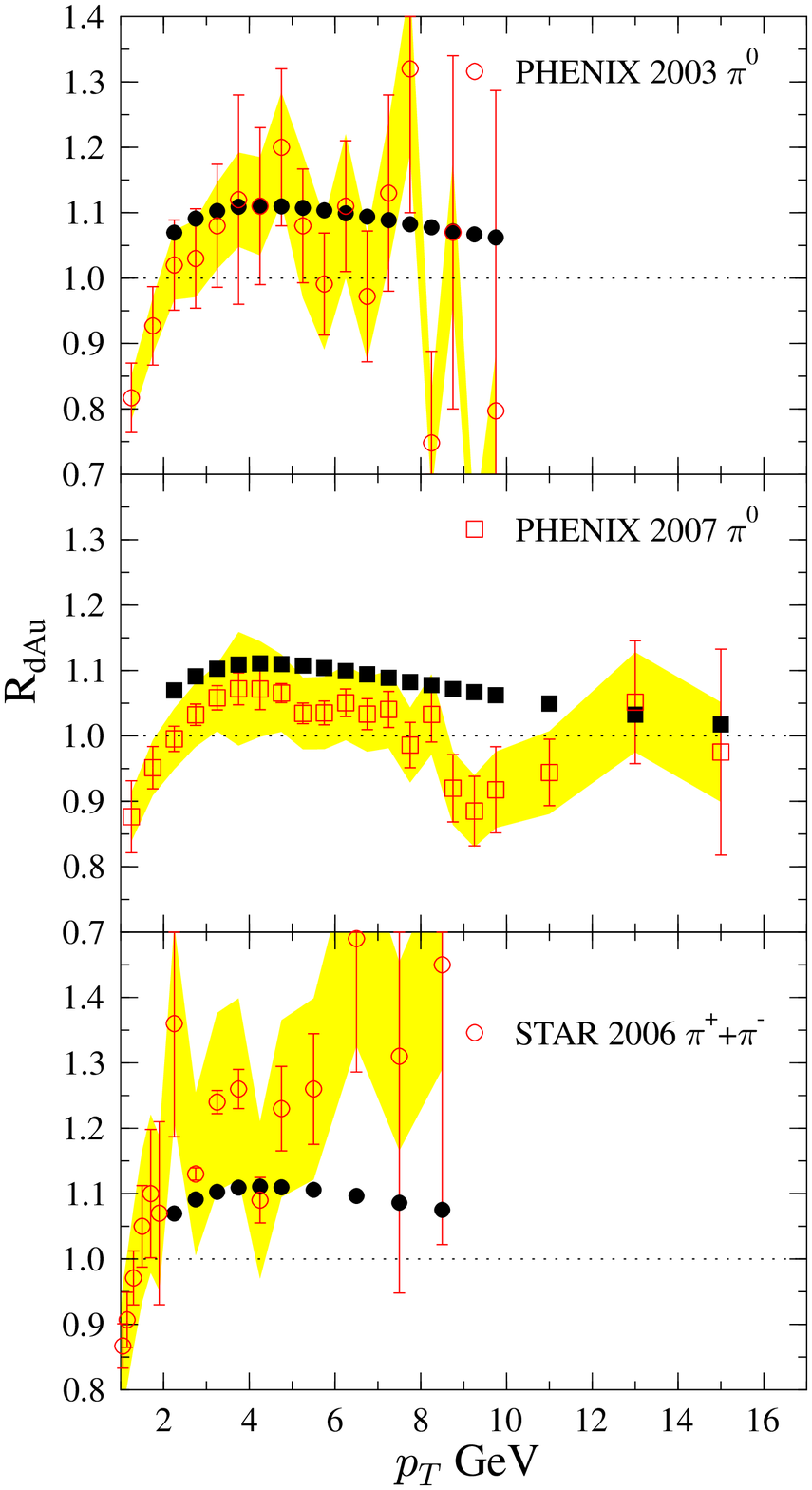}
\includegraphics[width=18pc]{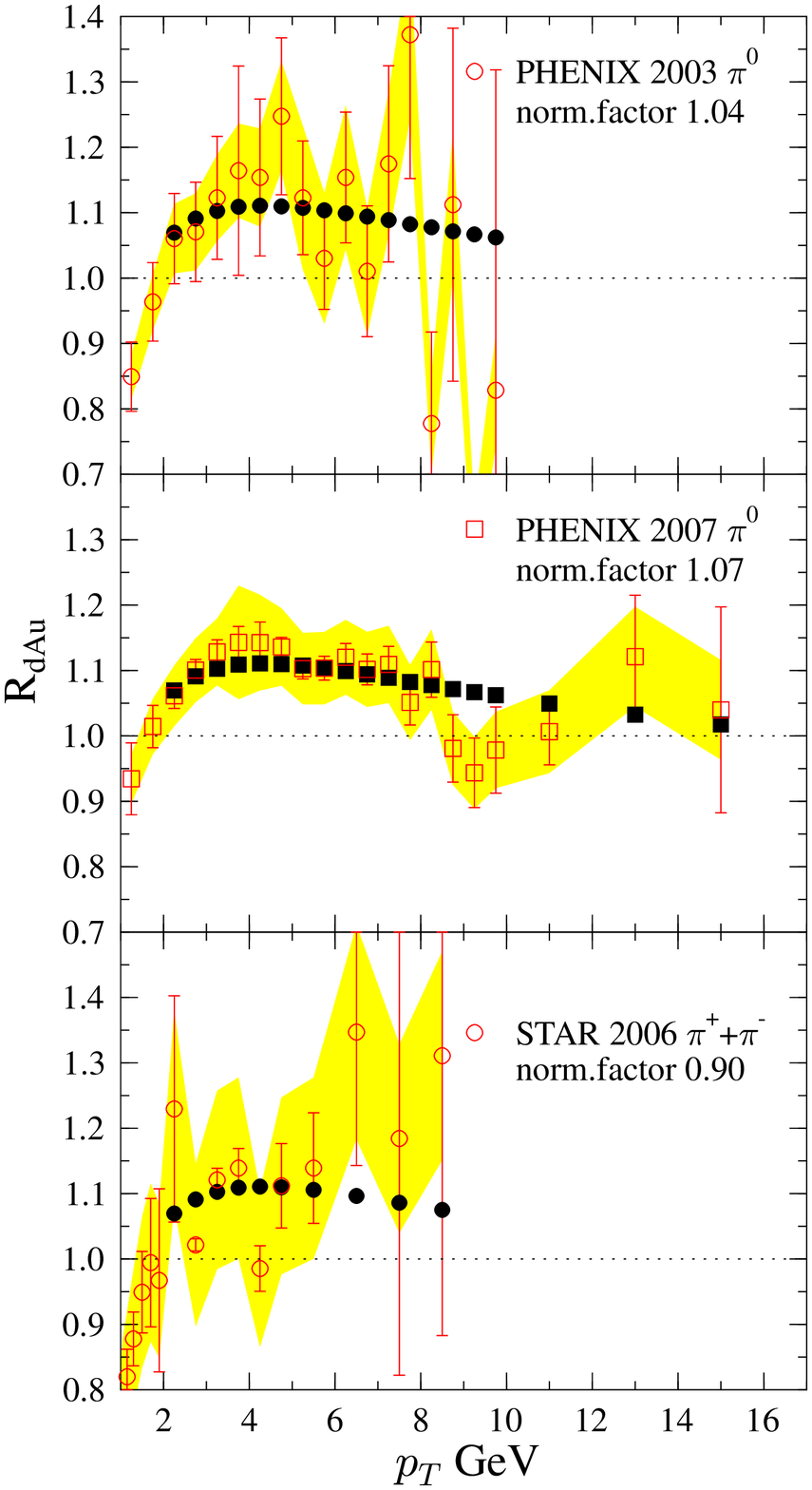}
\caption[]{\small 
The computed $R_{\rm dAu}$ (filled symbols) at midrapidity ($\eta=0$) for inclusive pion
production compared with the PHENIX \cite{Adler:2003ii,Adler:2006wg} and STAR \cite{Adams:2006nd} 
data (open symbols). The error bars are the statistical uncertainties, and the shaded bands indicate 
the point-to-point systematic errors. The additional overall normalization uncertainties are $10\%$ 
for the PHENIX data and $17\%$ for the STAR data. The left panels show the comparison without the 
normalization factor $f_N$. In the right panels, from top to bottom, the data have been multiplied by 
the optimized values $f_N = 1.04$, $f_N = 1.07$ and $f_N = 0.90$.}
\label{Fig:PHENIX_STAR}
\end{figure}

The comparison of our results with the experimental data used in the global fit is presented in Figs. 
\ref{Fig:RF2AC} -- \ref{Fig:RF2SnC} for DIS, in Figs. \ref{Fig:E772} -- \ref{Fig:E886} for DY, and
in Figs.~\ref{Fig:BRAHMS} -- \ref{Fig:PHENIX_STAR} for the RHIC data.
In all figures, the open (red) symbols denote the experimental data. The error bars for the DIS and 
DY cases correspond to the point-to-point statistical and systematic errors added in quadrature, 
while for the RHIC data the statistical and systematic errors are shown separately.

In comparison with our previous work, Ref.~\cite{Eskola:2007my}, the Copper and Lithium data in 
Figs.~\ref{Fig:RF2A1}-\ref{Fig:RF2Cu} have now been added into the analysis.
In  Fig.~\ref{Fig:RF2AC}, we see that shadowing for heavy nuclei (Sn and Pb) has now gotten stronger, 
and in Fig.~\ref{Fig:RF2SLAC} that we now reproduce the largest-$x$ DIS data better than before.
The smallest-$x$ panel of Fig. \ref{Fig:RF2SnC} is one of the key issues in this paper and the 
obtained  $Q^2$ slopes will be separately commented in Sec. \ref{sec:discussion} below.
Figure~\ref{Fig:E772} shows that for the DY ratios the $A$ systematics at the smallest values of 
$x_2$ have been improved: now also the Tungsten data are reproduced well. This improvement is 
reflected also in the large-$x_1$ part of the W/Be ratio in Fig.~\ref{Fig:E886}.

The data set that plays a major role in constraining the gluon modifications in the present analysis, 
is the inclusive negatively-charged hadron production at forward direction ($\eta = 2.2$ and $\eta = 
3.2$) measured by the BRAHMS collaboration at RHIC, shown in Fig.~\ref{Fig:BRAHMS}. For the data 
sample we include in the global fit, $p_T\ge 2$~GeV, the optimized normalization factor is close to 
one, $f_N=1.02$. 

Figure \ref{Fig:PHENIX_STAR} presents the comparison with the PHENIX and STAR measurements of 
inclusive pion production at midrapidity ($\eta \sim 0$). The need of a treatment which accounts for 
normalization uncertainties is clearly demonstrated by this figure. Although all data sets agree 
within the given large uncertainties, the general trend in the STAR data is somewhat different from 
the PHENIX data, as can be seen in the uncorrected case ($f_N=1$), shown on the left-hand side of the 
figure. Taking into account the normalization uncertainties as provided by the modified definition of 
$\chi^2$ in Eqs.~(\ref{eq:chi2mod_2}), a good fit with both PHENIX and STAR data sets becomes indeed 
possible -- see the right-hand side of Fig.~\ref{Fig:PHENIX_STAR}, where the optimized normalization 
factors for the PHENIX data are $f_N=1.04$ and 1.07 and for the STAR data $f_N=0.90$.

\section{Discussion}
\label{sec:discussion}

\begin{figure}[htbp]
\center
\includegraphics[scale=0.7]{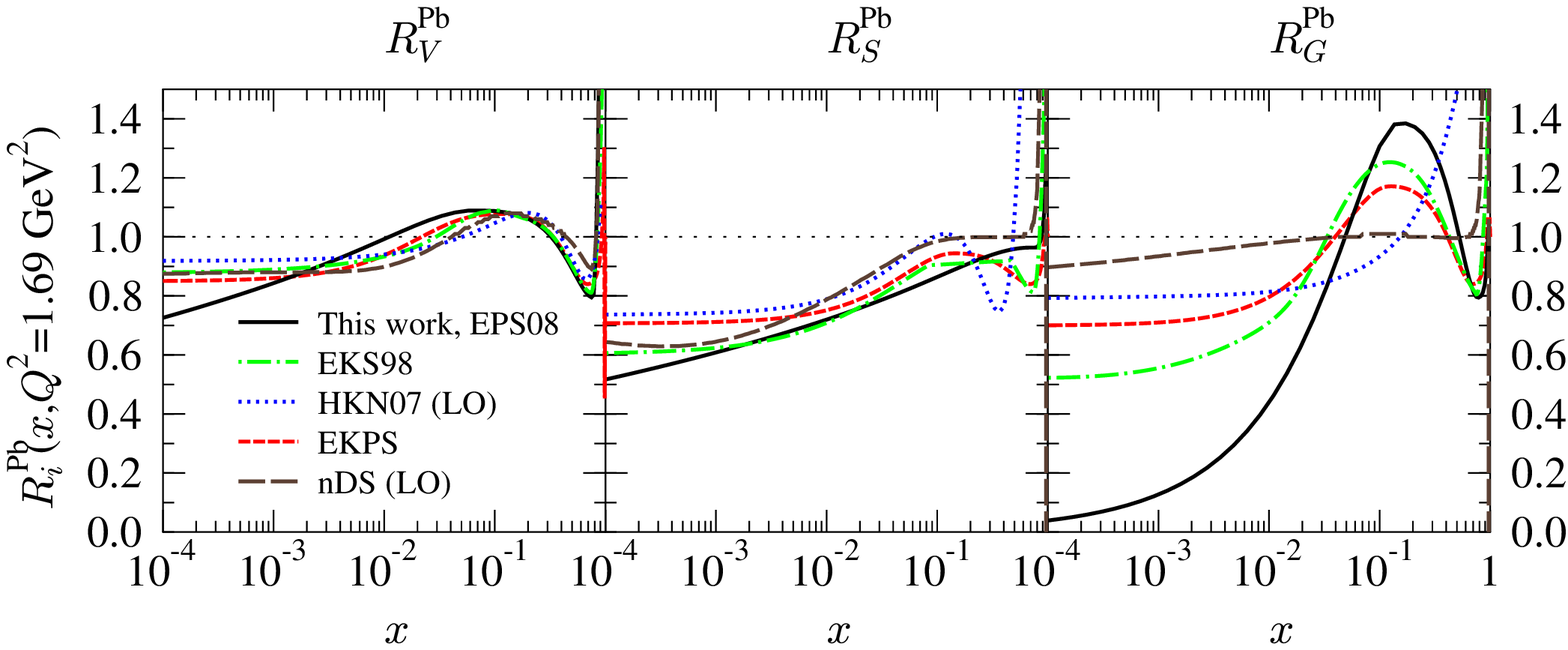}
\caption[]{\small Comparison of the average valence and sea quark, and gluon modifications 
at $Q^2 = 1.69$~GeV$^2$ for Pb nucleus from LO global DGLAP analyses EKS98~\cite{Eskola:1998df}, 
HKN07~\cite{Hirai:2007sx}, nDS~\cite{deFlorian:2003qf}, EKPS~\cite{Eskola:2007my} and this work 
EPS08.
}
\label{Fig:gluoncomp}
\end{figure}

Figure~\ref{Fig:gluoncomp} shows a comparison of the nuclear effects in the average valence quark, 
average sea quark and gluon distributions at our initial scale $Q_0^2=1.69$~GeV$^2$, as obtained 
for a Lead nucleus in the LO DGLAP analyses here (EPS08), in HKN07~\cite{Hirai:2007sx}, 
in nDS~\cite{deFlorian:2003qf},
and in our previous works EKPS~\cite{Eskola:2007my} and EKS98~\cite{Eskola:1998df}. The figure 
demonstrates the fact that while the average effects in the valence quarks and in the mid-$x$ region 
($0.01\lsim x\lsim 0.2$) of the sea quarks are relatively well under control, quite large 
uncertainties remain in the large-$x$ ($x\gsim 0.2$) sea quark and gluon distributions. In 
particular, the gluon shadowing which we obtain here on the basis of the BRAHMS data, is clearly 
stronger than in the previous global analyses.
A strong gluon shadowing has been suggested before at least in the Glauber-Gribov framework 
\cite{Frankfurt:2003zd,Tywoniuk:2007xy} (see also the review \cite{Armesto:2006ph}) and also in the 
context of DGLAP evolution \cite{Eskola:1992zb},
but not in a global DGLAP analysis where constraints from DIS, DY and RHIC hadron production data are 
simultaneously imposed.

\begin{figure}[htbp]
\center
\includegraphics[scale=0.63]{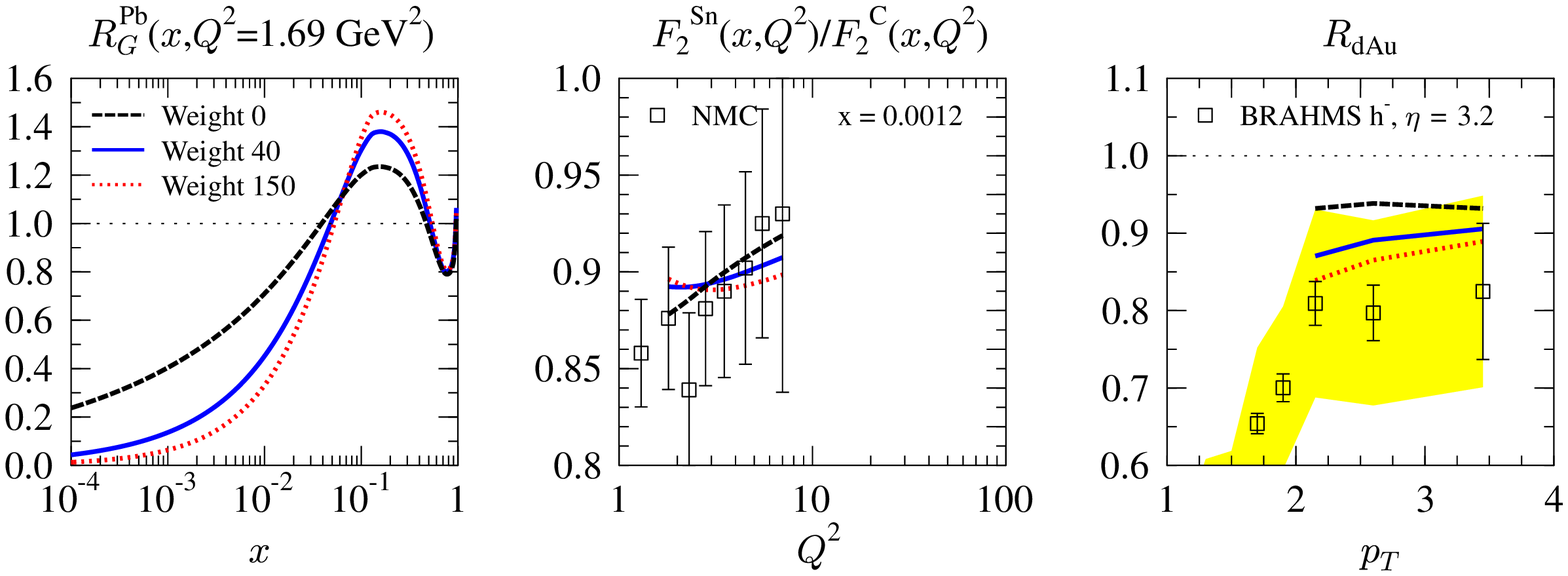}
\caption[]{\small The correlation between the gluon shadowing (left), the $Q^2$ slopes of $F_2^{\rm 
Sn}/F_2^{\rm C}$ (middle) and the forward-$\eta$ $R_{\rm dAu}$ (right) when weights $w_N=0$, $40$ and 
$150$ are assigned for the BRAHMS data. In the right panel, the comparison is shown
without an overall normalization factor, i.e. with $f_N=1$.
}
\label{Fig:weightcomp}
\end{figure}

The obtained gluon shadowing reflects a compromise between a weaker gluon shadowing 
suggested by the NMC 96 DIS data \cite{Arneodo:1995cs} (the first panels in Fig.~\ref{Fig:RF2SnC}), 
and a stronger effect demanded by the BRAHMS data. 
Regarding these constraints, we note the following:

First, the DIS data \cite{Arneodo:1995cs} show that the $Q^2$ dependence  of $R_{F_2}$ at 
$x\sim 0.01$ in the region $Q^2\gsim 1$~GeV$^2$ is very weak. Given this, the small-$x$ approximation 
of the DGLAP equations \cite{Prytz:1993vr,Eskola:2002us}, 
\begin{equation}
\frac{\partial R_{F_2}^{\rm A}(x,Q^2)}{\partial \log Q^2} \approx \frac{10\alpha_s}{27\pi} \frac{x 
g(2x,Q^2)}{F_2^{\rm D}(x,Q^2)} \times \left[ R_{G}^{\rm A}(2x,Q^2) - R_{F_2}^{\rm A}(x,Q^2)\right],
\label{eqn:Q2slope}
\end{equation}
then indicates that gluon shadowing is restricted to be similar to what has been measured for $F_2$. 

Second, using Eq.~(\ref{eqn:Q2slope}) above, and the fact that 
the $\log Q^2$-slope of $F_2^{\rm Sn}/F_2^{\rm C}$ measured by NMC has been observed to be positive 
at $x=0.125$ -- see the first panel in Fig.~\ref{Fig:RF2SnC} -- we deduce that 
\begin{equation}
\left. \frac{R_G^{\rm Sn}(2x,Q^2)}{R_G^{\rm C}(2x,Q^2)} \right|_{x \approx 0.0125} >
\left. \frac{R_{F_2}^{\rm Sn}(x,Q^2)}{R_{F_2}^{\rm C}(x,Q^2)}\right|_{x \approx 0.0125.}
\label{eqn:slope2}
\end{equation}
This indicates that the $A$ dependence of gluon shadowing is weaker than that of $F_2$. 

Third, the $A$ dependence of gluon shadowing must still be strong enough in order to reproduce the 
BRAHMS data. The flattening of the $Q^2$-slope seen in the first panel of  Fig.~\ref{Fig:RF2SnC} 
indicates that the gluon shadowing now obtained -- the $A$ dependence of the gluon modifications in 
particular -- is already so strong that it is in the brink of violating the condition 
(\ref{eqn:slope2}). In this sense the gluon shadowing in the present global fit is the strongest 
possible one which is still in agreement with the DIS data. 

Fourth, a balance between the constraints that the NMC 96 and BRAHMS data offer for the gluon 
shadowing, is obtained by assigning suitable relative weights, see Table~\ref{Table:Data}. 
Since these data sets drive the fit to opposite directions, 
the resulting gluon shadowing obviously depends on the weights introduced. 
To demonstrate this sensitivity, we have repeated the analysis by varying the BRAHMS data weights as 
follows: By setting $w_N=0$, we remove this data set from the analysis. Alternatively, by assigning a 
very large weight, $w_N=150$, to the BRAHMS data, we clearly overemphasize its importance. In both 
cases, the smallest-$x$ NMC 96 data set weight is kept unchanged ($w_N=10$, see 
Table~\ref{Table:Data}). The overall fits obtained in these extreme cases remain very good, giving 
$\chi^2/N=0.72$ and 0.73, correspondingly. Figure ~\ref{Fig:weightcomp} (left panel) shows the 
resulting gluon modifications in each case, along with a comparison to the smallest-$x$ NMC 96 data 
(middle panel) and the BRAHMS data (right panel). 
The figure clearly demonstrates how adding more weight to the BRAHMS data will eventually flip the 
sign of the computed $Q^2$ slopes of $F_2^{\rm Sn}/F_2^{\rm C}$ --- a phenomenon which on the basis 
of the systematics seen in the NMC 96 data would be an unwanted feature. With a weight factor 
$w_N=40$ for the BRAHMS data
(making the effective number of the BRAHMS data points the same as in the three smallest-$x$ panels 
of the NMC 96 data), we reach the strongest possible gluon shadowing, and thus a fair agreement with 
the measured forward-rapidity $R_{\rm dAu}$,  without such a sign flip.

Naively, in the RHIC hadron data, we could well expect that hadrons at fixed $\eta$ and $p_T$ would 
dominantly come from partons of higher transverse momenta, $q_T\sim 1.5...2p_T$ and the same rapidity 
$\eta$ \cite{Eskola:2002kv}, whose production would mainly probe the nPDFs at momentum fractions 
$x_2= \frac{q_T}{\sqrt{s}}({\rm e}^{-\eta}+{\rm e}^{-y_2})\approx \frac{4p_T}{\sqrt{s}} {\rm 
e}^{-\eta}\sim 10^{-3}$, assuming $2\rightarrow 2$ parton production kinematics, and taking $p_T\sim 
2$~GeV and $y_2\sim\eta=3.2$. We notice, however, that the ratio $R_{\rm dAu}$ at  $p_T=2$~GeV in 
Fig.~\ref{Fig:BRAHMS} is considerably larger than the gluon shadowing we obtain at these values of 
$x$, see Fig.~\ref{Fig:gluoncomp} (solid line). This is due to two reasons:
First, as shown in Fig. 13 of \cite{Eskola:2007my}, the DGLAP evolution from $Q_0$ to 
the few-GeV region increases the ratio $R_G^A$ substantially. Second, 
the integration over the partonic $q_T$ and over the unobserved parton rapidity $y_2$
causes a significant smearing of the $x$-range probed, especially towards larger $x$ (see also 
Ref.~\cite{Guzey:2004zp} and Table 1 in \cite{Vogt:2004hf}). Thus, the ratio $R_{\rm dAu}$ at forward 
$\eta$ is in fact sensitive not only to nuclear gluon shadowing but also to gluon antishadowing -- 
and antishadowing in turn amplifies
when shadowing gets stronger. Therefore, even a large change in gluon shadowing induces only a 
moderate change in the computed ratio $R_{\rm dAu}$, and a significant gluon shadowing is required in 
order to reproduce the BRAHMS data at $p_T\ge 2$~GeV.

As seen in Figs.~\ref{Fig:BRAHMS} and \ref{Fig:PHENIX_STAR}, we have a good fit of the nuclear 
modification factor $R_{\rm dAu}$ at $p_T\ge 2$~GeV. We cannot, however, reduce $R_{\rm dAu}$ by 
strengthening the gluon shadowing as much as the BRAHMS data below 2 GeV would require
without violating the DIS data constraints. This is also one of the reasons for excluding 
the region $p_T < 2$ GeV of the RHIC data from this analysis. As explained above, we also are 
hesitant to push the nuclear case too far into the small-$p_T$ region, for we cannot reproduce the 
shape of the absolute $p_T$ spectra in p+p collisions well enough there, and for we do not consider 
impact-parameter dependence of nPDFs or a more detailed centrality selection here.

As illustrated by Fig.~\ref{Fig:gluoncomp}, a saturation of shadowing at $x\rightarrow 0$ was assumed 
in previous global analyses. This assumption is now relaxed with the aim to study the strongest gluon 
shadowing allowed by present experimental data. It is worth emphasizing that the behavior of the 
nuclear corrections at the smallest values of $x$ ($x\lsim 10^{-3}$) is still largely determined by 
the assumed shape of the fit functions in Eq.~(\ref{eq:R}). This limitation is common to any global 
fit (for nPDFs as well as for the free proton PDFs) in those regions of phase space which are poorly 
or not at all constrained by the data.

\section{Conclusions}
\label{sec:conclusions}

We have improved the global analysis of nPDFs in two important ways:
First, by taking the RHIC data into account in such analysis for the first time,
we have extended the constrained $x$ region down to $x\gsim 10^{-3}$. Second, we have improved the 
$\chi^2$ minimization procedure by introducing weighting of different data sets and by explicitly 
accounting for the overall normalization errors quoted by the experiments.

One of the main goals of this paper is to study to what extent a strong gluon shadowing suggested by 
the BRAHMS data can be accommodated together with the DIS and DY data in a global analysis. 
We conclude that a simultaneous fit of DIS, DY and high-$p_T$ ($p_T\ge 2$ GeV) hadron production data 
from RHIC  at forward rapidities (BRAHMS negative hadrons), is indeed possible within the DGLAP 
framework without invoking any new suppression mechanism. Thanks to the improved treatment of data 
normalization errors, we obtain also a good agreement with the RHIC pion data at mid-rapidity (STAR 
and PHENIX). The very good quality of the global fit obtained suggests that a well-working universal 
set of nPDFs can be extracted in the framework of collinear factorization. Within an improved global 
$\chi^2$ analysis, and with an emphasis on the RHIC forward-rapidity data, we obtain a stronger gluon 
shadowing than in the previous global nPDF analyses, see Fig.~\ref{Fig:gluoncomp}. These are the main 
new results of this paper. The LO nPDF set we have obtained (EPS08), \emph{i.e.} a parametrization of 
the $x$, $Q^2$ and $A$-dependent nuclear modifications relative to CTEQ6L1, is available at 
\cite{nPDF_address} 
for practical use.

As discussed above, the amount of gluon shadowing obtained depends on the weight assigned to the 
BRAHMS data, and equally good overall fits can be obtained also when the weights are smaller and the 
resulting gluon shadowing is weaker. Until more data become available to resolve this problem, the 
nuclear gluon distributions suffer from considerable uncertainties. To estimate the effects of these 
uncertainties in the 
hard process cross sections one computes in LO, we recommend to use the current results, EPS08, in 
parallel with the previous LO results EKS98 \cite{Eskola:1998df}, nDS~\cite{deFlorian:2003qf} and 
HKN07 \cite{Hirai:2007sx}. 

Regarding inclusive hadron production in nuclear and hadronic collisions, computed here in the 
collinear factorization framework, we would like to emphasize that we have limited the study to the 
region $p_T\ge 2$~GeV:  
Within the present global analysis, we cannot reproduce the sudden drop of the ratio $R_{\rm dAu}$ 
measured for negative hadrons by BRAHMS at $p_T<2$~GeV at forward rapidities, see 
Fig.~\ref{Fig:BRAHMS}, or the very strong suppression of $R_{\rm dAu}$ measured by STAR for $\pi^0$ 
at $\eta=4$ and $p_T<2$ GeV \cite{Adams:2006uz} -- a yet stronger gluon shadowing needed for this 
would clearly lead into a contradiction with the $\log Q^2$ slopes of $F_2^{\rm Sn}/F_2^{\rm C}$ 
measured at DIS. More detailed work on fragmentation functions,  impact parameter dependence of the 
nPDFs as well as further developments in the fit functions is required in order to make firmer 
conclusions on the applicability of the DGLAP-evolved universal nPDFs in this region. 
Regarding the fragmentation functions, we anticipate that considering a more detailed charged 
separation (see e.g. \cite{de Florian:2007hc,deFlorian:2007aj,Albino:2008fy}) in hadron production 
would tend to increase the computed $R_{\rm dAu}$
rather than decrease it. Such further complication in extracting the gluon shadowing
from the BRAHMS data is, however, not considered here, since the inclusion of the charge separation 
becomes more reliable only in NLO. 

Our next goal is to perform this analysis in NLO, as well as, when the data become finalized, include 
other RHIC data sets, such as photon production in d+Au and Au+Au from PHENIX, into the analysis. In 
general, any further constraints for the gluon distributions are more than welcome. For example, the 
ratio $R_{\rm dAu}$ for $D$ mesons to be (hopefully soon) measured at RHIC will be extremely useful, 
at any rapidity. The cleanest environment for the nPDFs measurements would be in the DIS experiments 
at eRHIC \cite{Cazaroto:2008qh} and LHeC colliders now being discussed. Before the possible 
realization of these machines, we hope that the present study in its part demonstrates how important 
it would be for the correct determination of universal nPDFs to have a systematic proton-nucleus 
program also at the LHC: further constraints for nuclear gluons in the yet unexplored regions of the 
$x,Q^2$ plane are absolutely necessary for understanding QCD parton dynamics in high-energy nuclear 
and hadronic collisions.

\section*{Acknowledgements}
We thank the Academy of Finland, Projects 206024 and 115262, for financial support.
CAS is supported by Ministerio de Educaci\'on y Ciencia of Spain under a Ram\'on y Cajal contract.


\begin{thebibliography}{99}




\bibitem{Eskola:1998iy}
  K.~J.~Eskola, V.~J.~Kolhinen and P.~V.~Ruuskanen,
  Nucl.\ Phys.\ B {\bf 535} (1998) 351
  [arXiv:hep-ph/9802350].

\bibitem{Eskola:1998df}
  K.~J.~Eskola, V.~J.~Kolhinen and C.~A.~Salgado,
  Eur.\ Phys.\ J.\ C {\bf 9} (1999) 61
  [arXiv:hep-ph/9807297].

\bibitem{Eskola:2007my}
  K.~J.~Eskola, V.~J.~Kolhinen, H.~Paukkunen and C.~A.~Salgado,
  JHEP {\bf 0705} (2007) 002
  [arXiv:hep-ph/0703104].

\bibitem{Hirai:2001np}
  M.~Hirai, S.~Kumano and M.~Miyama,
  Phys.\ Rev.\ D {\bf 64} (2001) 034003
  [arXiv:hep-ph/0103208].

\bibitem{Hirai:2004wq}
  M.~Hirai, S.~Kumano and T.~H.~Nagai,
  Phys.\ Rev.\ C {\bf 70} (2004) 044905
  [arXiv:hep-ph/0404093].

\bibitem{deFlorian:2003qf}
  D.~de Florian and R.~Sassot,
  Phys.\ Rev.\ D {\bf 69} (2004) 074028
  [arXiv:hep-ph/0311227].

\bibitem{Hirai:2007sx}
  M.~Hirai, S.~Kumano and T.~H.~Nagai,
  arXiv:0709.3038 [hep-ph].

\bibitem{Schienbein:2007fs}
  I.~Schienbein, J.~Y.~Yu, C.~Keppel, J.~G.~Morfin, F.~Olness and J.~F.~Owens,
  arXiv:0710.4897 [hep-ph].

  

\bibitem{DGLAP} 
Y.~L.~Dokshitzer,
Perturbation Theory In Quantum 
Sov.\ Phys.\ JETP {\bf 46} (1977) 641
[Zh.\ Eksp.\ Teor.\ Fiz.\  {\bf 73} (1977) 1216];
V.~N.~Gribov and L.~N.~Lipatov,
Yad.\ Fiz.\  {\bf 15} (1972) 781
[Sov.\ J.\ Nucl.\ Phys.\  {\bf 15} (1972) 438];
V.~N.~Gribov and L.~N.~Lipatov,
Yad.\ Fiz.\  {\bf 15} (1972) 1218
[Sov.\ J.\ Nucl.\ Phys.\  {\bf 15} (1972) 675];
G.~Altarelli and G.~Parisi,
Nucl.\ Phys.\ B {\bf 126} (1977) 298.


  \bibitem{Vogt:2004hf}
  R.~Vogt,
  Phys.\ Rev.\  C {\bf 70} (2004) 064902.

\bibitem{Arsene:2004ux}
  I.~Arsene {\it et al.}  [BRAHMS Collaboration],
  Phys.\ Rev.\ Lett.\  {\bf 93} (2004) 242303
  [arXiv:nucl-ex/0403005].

\bibitem{satur}
  R.~Baier, A.~Kovner and U.~A.~Wiedemann,
  Phys.\ Rev.\ D {\bf 68}, 054009 (2003);
  D.~Kharzeev, Y.~V.~Kovchegov and K.~Tuchin,
  Phys.\ Rev.\ D {\bf 68} (2003) 094013;
  J.~L.~Albacete, N.~Armesto, A.~Kovner, C.~A.~Salgado and U.~A.~Wiedemann,
  Phys.\ Rev.\ Lett.\  {\bf 92}, 082001 (2004).

\bibitem{Stump:2001gu}
  D.~Stump {\it et al.},
  Phys.\ Rev.\  D {\bf 65} (2002) 014012
  [arXiv:hep-ph/0101051].

\bibitem{Adler:2003ii}
  S.~S.~Adler {\it et al.}  [PHENIX Collaboration],
  Phys.\ Rev.\ Lett.\  {\bf 91} (2003) 072303
  [arXiv:nucl-ex/0306021].

\bibitem{Adler:2006wg}
  S.~S.~Adler {\it et al.}  [PHENIX Collaboration],
  Phys.\ Rev.\ Lett.\  {\bf 98} (2007) 172302
  [arXiv:nucl-ex/0610036].

\bibitem{Adams:2006nd}
  J.~Adams {\it et al.}  [STAR Collaboration],
  Phys.\ Lett.\  B {\bf 637} (2006) 161
  [arXiv:nucl-ex/0601033].

\bibitem{nPDF_address}
http://www.jyu.fi/science/laitokset/fysiikka/en/research/highenergy/urhic/nPDFs

http://www-fp.usc.es/~phenom/nPDFs. 

\bibitem{Pumplin:2002vw}
  J.~Pumplin, D.~R.~Stump, J.~Huston, H.~L.~Lai, P.~Nadolsky and W.~K.~Tung,
  JHEP {\bf 0207} (2002) 012
  [arXiv:hep-ph/0201195].

  \bibitem{Eskola:1991ec}
  K.~J.~Eskola,
  Z.\ Phys.\  C {\bf 51} (1991) 633.

  \bibitem{Santorelli:1998yt}
  P.~Santorelli and E.~Scrimieri,
  Phys.\ Lett.\  B {\bf 459} (1999) 599
  [arXiv:hep-ph/9807572].

\bibitem{Kniehl:2000fe}
  B.~A.~Kniehl, G.~Kramer and B.~Potter,
  Nucl.\ Phys.\  B {\bf 582} (2000) 514
  [arXiv:hep-ph/0010289].

\bibitem{Eskola:2002kv}
  K.~J.~Eskola and H.~Honkanen,
  Nucl.\ Phys.\  A {\bf 713} (2003) 167
  [arXiv:hep-ph/0205048].

  \bibitem{Borzumati:1995ib}
  F.~Borzumati and G.~Kramer,
  Z.\ Phys.\  C {\bf 67} (1995) 137
  [arXiv:hep-ph/9502280].

  \bibitem{Aurenche:1999nz}
  P.~Aurenche, M.~Fontannaz, J.~P.~Guillet, B.~A.~Kniehl and M.~Werlen,
  Eur.\ Phys.\ J.\  C {\bf 13} (2000) 347
  [arXiv:hep-ph/9910252].


\bibitem{Gomez:1993ri}
  J.~Gomez {\it et al.},
  Phys.\ Rev.\ D {\bf 49} (1994) 4348.

  \bibitem{Amaudruz:1995tq}
  P.~Amaudruz {\it et al.}  [New Muon Collaboration],
  Nucl.\ Phys.\  B {\bf 441} (1995) 3
  [arXiv:hep-ph/9503291].

\bibitem{Arneodo:1995cs}
  M.~Arneodo {\it et al.}  [New Muon Collaboration.],
  Nucl.\ Phys.\ B {\bf 441} (1995) 12
  [arXiv:hep-ex/9504002].

\bibitem{Arneodo:1996rv}
  M.~Arneodo {\it et al.}  [New Muon Collaboration],
  Nucl.\ Phys.\ B {\bf 481} (1996) 3.

\bibitem{Alde:1990im}
  D.~M.~Alde {\it et al.},
  Phys.\ Rev.\ Lett.\  {\bf 64} (1990) 2479.

\bibitem{Vasilev:1999fa}
  M.~A.~Vasilev {\it et al.}  [FNAL E866 Collaboration],
  Phys.\ Rev.\ Lett.\  {\bf 83} (1999) 2304
  [arXiv:hep-ex/9906010].

\bibitem{Ashman:1992kv}
  J.~Ashman {\it et al.}  [European Muon Collaboration],
  Z.\ Phys.\  C {\bf 57} (1993) 211.
 
\bibitem{Arneodo:1996ru}
  M.~Arneodo {\it et al.}  [New Muon Collaboration],
  Nucl.\ Phys.\ B {\bf 481} (1996) 23.


  \bibitem{Tannenbaum:2007sy}
  M.~J.~Tannenbaum,
  arXiv:0707.1706 [nucl-ex].

\bibitem{James:1975dr}
  F.~James and M.~Roos,
  Comput.\ Phys.\ Commun.\  {\bf 10} (1975) 343.

  \bibitem{Frankfurt:2003zd}
  L.~Frankfurt, V.~Guzey and M.~Strikman,
  Phys.\ Rev.\  D {\bf 71} (2005) 054001
  [arXiv:hep-ph/0303022].

\bibitem{Tywoniuk:2007xy}
  K.~Tywoniuk, I.~Arsene, L.~Bravina, A.~Kaidalov and E.~Zabrodin,
  Phys.\ Lett.\  B {\bf 657} (2007) 170
  [arXiv:0705.1596 [hep-ph]].

  \bibitem{Armesto:2006ph}
  N.~Armesto,
  J.\ Phys.\ G {\bf 32} (2006) R367
  [arXiv:hep-ph/0604108].
  
  \bibitem{Eskola:1992zb}
  K.~J.~Eskola,
  Nucl.\ Phys.\  B {\bf 400} (1993) 240.

\bibitem{Prytz:1993vr}
  K.~Prytz,
  Phys.\ Lett.\  B {\bf 311} (1993) 286.

\bibitem{Eskola:2002us}
  K.~J.~Eskola, H.~Honkanen, V.~J.~Kolhinen and C.~A.~Salgado,
  Phys.\ Lett.\  B {\bf 532} (2002) 222
  [arXiv:hep-ph/0201256].

\bibitem{Guzey:2004zp}
  V.~Guzey, M.~Strikman and W.~Vogelsang,
  Phys.\ Lett.\  B {\bf 603} (2004) 173
  [arXiv:hep-ph/0407201].
  
\bibitem{Adams:2006uz}
  J.~Adams {\it et al.}  [STAR Collaboration],
  Phys.\ Rev.\ Lett.\  {\bf 97} (2006) 152302
  [arXiv:nucl-ex/0602011].
  
  \bibitem{de Florian:2007hc}
  D.~de Florian, R.~Sassot and M.~Stratmann,
  Phys.\ Rev.\  D {\bf 76} (2007) 074033
  [arXiv:0707.1506 [hep-ph]].
  
  \bibitem{deFlorian:2007aj}
  D.~de Florian, R.~Sassot and M.~Stratmann,
  Phys.\ Rev.\  D {\bf 75} (2007) 114010
  [arXiv:hep-ph/0703242].
  
  \bibitem{Albino:2008fy}
  S.~Albino, B.~A.~Kniehl and G.~Kramer,
  arXiv:0803.2768 [hep-ph].

  \bibitem{Cazaroto:2008qh}
  E.~R.~Cazaroto, F.~Carvalho, V.~P.~Goncalves and F.~S.~Navarra,
  arXiv:0804.2507 [hep-ph].

\end{thebibliography}
\end{document}